\documentclass[twocolumn,secnumarabic,amssymb, nobibnotes, aps, prd,superscriptaddress]{revtex4-2}

\usepackage{graphicx}
\usepackage{color}
\usepackage{amsmath}
\usepackage{gensymb}
\usepackage{tabularx}
\usepackage{multirow}
\usepackage{notes2bib}
\usepackage{ulem}
\usepackage[left,modulo]{lineno}
\usepackage{upgreek}

\makeatletter
\newcommand{\maintextlabel}[2]{\def\@currentlabel{#2}\label{#1}}
\makeatother

\begin{document}

\onecolumngrid
This article may be downloaded for personal use only. Any other use requires prior permission of the author and AIP Publishing. This article appeared in Marceau Hénot, François Ladieu; Non-linear physical aging of supercooled glycerol induced by large upward ideal temperature steps monitored through cooling experiments. J. Chem. Phys. 9 June 2023; 158 (22): 224504. and may be found at \url{https://doi.org/10.1063/5.0151518}.

\title{Non-linear physical aging of supercooled glycerol induced by large upward ideal temperature steps monitored through cooling experiments}

\author{Marceau H\'enot}
\email[Corresponding author: ]{marceau.henot@cea.fr}
\affiliation{SPEC, CEA, CNRS, Université Paris-Saclay, CEA Saclay Bat 772, 91191 Gif-sur-Yvette Cedex, France.}

\author{François Ladieu}
\affiliation{SPEC, CEA, CNRS, Université Paris-Saclay, CEA Saclay Bat 772, 91191 Gif-sur-Yvette Cedex, France.}

\date{\today}
\begin{abstract}
The physical aging of supercooled glycerol induced by upward temperature steps of amplitude reaching 45~K was studied by a new method consisting in heating a micrometer-thick liquid film at a rate of up to 60~000 ~K/s, holding it at a constant high temperature for a controlled duration before letting it quickly cool down to the initial temperature. By monitoring the final slow relaxation of the dielectric loss, we were able to obtain quantitative information on the liquid response to the initial upward step. The so-called TNM formalism provided a good description of our observations despite the large distance from equilibrium provided that different values of the nonlinearity parameter were used for the cooling phase and for the (much further from equilibrium) heating phase. In this form, it allowed to precisely quantify how to design an ideal temperature step, \textit{i.e.} where no relaxation occurs during the heating phase. It helped bringing a clear physical understanding of how the (kilosecond long) final relaxation is related to the (millisecond long) liquid response to the upward step. Finally, it made possible the reconstruction of the fictive temperature evolution immediately following a step, evidencing the highly non-linear character of the liquid response to such large amplitude temperature steps. This work illustrates both the strengths and limitations of the TNM approach. This new experimental device offers a promising tool to study far-from-equilibrium supercooled liquids through their dielectric response.

\end{abstract}
\maketitle

\section{Introduction.}
Physical aging corresponds to the evolution of out-of-equilibrium material properties caused by structural rearrangements. This phenomenon is observed in various disordered systems, from network-forming materials~\cite{tool1946relation, narayanaswamy1971model, micoulaut2016relaxation} to supercooled liquids~\cite{leheny_frequency-domain_1998, lunkenheimer_glassy_2005, brun2012evidence, hecksher2010physical, hecksher2015communication, roed2019generalized, riechers_predicting_2022}, amorphous polymer~\cite{kovacs1958contraction, mckenna1995physics}, colloids~\cite{abou2001aging}, granular materials~\cite{richard2005slow} and even active glasses~\cite{mandal2020multiple}. For glassy materials, which are in practice forever stuck in an out-of-equilibrium state, physical aging can have important consequences on long-term material performances such as mechanical strength~\cite{andersen2019accelerated} or optical properties~\cite{jena2019study}. This phenomenon also offers a playground for theories that aim for a better fundamental understanding of the glassy state~\cite{arceri2020glasses}. 

Physical aging can be studied experimentally by setting a system out-of-equilibrium through a controlled modification of its external conditions (temperature, pressure, electric field, etc) and measuring the relaxation of one of its macroscopic properties (refractive index, enthalpy, dielectric loss, etc) toward an equilibrium state. The simplest kind of experiment is the ideal temperature step (or jump): the temperature of the system is varied from $T_0$ to $T_0 + \Delta T$ quickly enough so that just after the step, the system has not yet started to relax. For very small step amplitudes, re-equilibrating the system takes a typical aging time which is simply related to the mean structural relaxation time $\tau_\alpha(T_0)$ (although the relaxation can be non-exponential). For larger steps, $\tau_\alpha$ strongly varies between $T_0$ and $T_0+\Delta T$ making the response non-linear. Added to this is the so-called fictive temperature effect in which, for the same final temperature, the response to a downward temperature step is faster, due to auto-acceleration, than an upward step which is subject to auto-retardation~\cite{lillie1933viscosity, kovacs1964transition, hecksher2010physical, riechers_predicting_2022}. The TNM formalism (for Tool-Narayanaswamy-Moynihan)~\cite{tool1946relation, narayanaswamy1971model, moynihan1976structural} offers an elegant way of interpreting physical aging by introducing the notion of \textit{material time} $\xi$: a dimensionless quantity that measures the passing of time with respect to the instantaneous relaxation rate of the material $\gamma$. The link with the external observer time $t$ being $\mathrm{d}\xi = \gamma \mathrm{d}t$. From the viewpoint of the material time, the system, characterized by its fictive temperature $T_\mathrm{f}$ (that quantifies, in temperature unit, the distance from equilibrium~\cite{mauro2009fictive}), responds linearly to any external temperature variation $T(\xi)$: %
\begin{equation}
   T_\mathrm{f}(\xi) = T(\xi) - \int_0^\xi M(\xi - \xi^\prime)\frac{\mathrm{d}T}{\mathrm{d}\xi^\prime}\mathrm{d}\xi^\prime
\label{eq_TNM_resp}
\end{equation}
where $M$ is a memory kernel that is assumed to depend only on the nature of the system but not on its temperature. The response appears non-linear only from the viewpoint of an external observer whose time flows, from the material perspective, at a $T_\mathrm{f}$ dependent rate. Recently, the existence of an internal clock linking the structural and dielectric processes was demonstrated~\cite{hecksher2010physical} and the TNM model was successively applied for various molecular liquids for ideal temperature step experiments with amplitude ranging from 50~mK to 8~K~\cite{hecksher2015communication, roed2019generalized, riechers_predicting_2022}. Other models were used in the literature to interpret aging experiments such as the KAHR approach~\cite{kovacs1979isobaric} which takes into account the temporally heterogeneous nature of the relaxation but was shown to give equivalent prediction to the TNM model~\cite{mazinani2012enthalpy}. Lunkenheimer~\textit{et al.}~\cite{lunkenheimer_glassy_2005} used a much simpler approach, with only one adjustable parameter, leading to an equally good agreement with experiments~\cite{richert_derivation_2013} but limited to ideal temperature steps.

Recently, ultrastable organic glasses (consisting of thin vapour-deposited films) offered the possibility to study the far from equilibrium response of low $T_\mathrm{f}$ systems annealed at significantly higher (up to $+60$~K) 
temperature. In this case, the behavior appears heterogeneous with liquid patches growing inside the glass~\cite{sepulveda2014role, vila2020nucleation}. This process competes with the more classic progressive rejuvenation that dominates for smaller temperature step amplitudes~\cite{vila2023emergence}.

In this article, we investigate the response of supercooled glycerol to upward temperature steps of high amplitude, ranging from 34 to 45~K, with a heating time between 0.7 and 6~ms. The liquid response is measured by cooling the liquid back to its starting temperature after a controlled time spent at high temperature and studying the slow relaxation of the dielectric loss. We first present the new experimental setup developed to apply such large temperature step amplitudes and heating rates and we show how the duration of the high temperature phase affects the aging curves. The TNM formalism is applied and the necessary adjustment required to obtain a good description of the experimental observations, despite the far reached distance from equilibrium, is discussed. This illustrates the limitation and potential of this approach, which is then exploited to get a physical understanding of how measuring the final relaxation provides information about the millisecond long response of the liquid to upward temperature steps. From this, the necessary and sufficient condition on the step amplitude and the heating rate needed to reach the ideal step regime are evaluated and discussed. Finally, we use the model to reconstruct the liquid response from our measurements and illustrate the highly non-linear behavior of the system to such large variations in temperature. While this method does not allow to determine unequivocally the heterogeneous or homogeneous nature of the liquid response, it is a promising tool to study the far-from-equilibrium response of supercooled liquids. 

\section{Methods.}
\begin{figure}[htbp]
\centering
\includegraphics[width=\columnwidth]{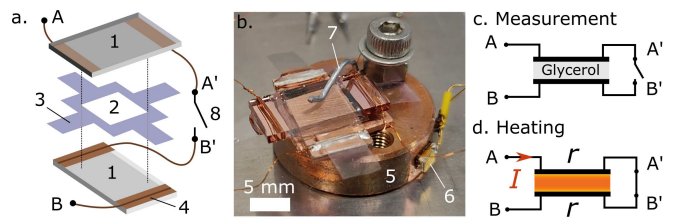}
 \caption{(a) Exploded view of the experimental setup which consists of two resistive electrodes (1) sandwiching a glycerol layer (2) of thickness $h$ and a PET spacer (3). Each electrode is connected at both ends by copper layers (4) to wires (A-A' and B-B'). (b) Picture showing the electrodes attached by a spring (7) on a copper frame (5) in contact with a thermometer (6). A switch (8) is used to select between two modes: a measurement mode (c) allows to probe the dielectric properties of the glycerol and a heating mode (d) in which an electric current $I$ flows through the electrodes of resistance $r$, producing a thermal flux heating the glycerol.}
\label{fig1}
\end{figure}
\textbf{Experimental setup.}  The setup is presented in fig.~\ref{fig1}a and b and described in more detail in the supplementary. A layer of glycerol (purchased from Sigma-Aldrich, $\geq 99.5$~\%) of thickness $h=15\pm 1$~µm is sandwiched between two electrodes separated by a PET spacer. The electrodes consist of glass plates covered by a conductive layer of indium tin oxide (ITO) that each has an electrical resistance $r=47\pm 2~\Omega$ between their two ends. The ensemble is placed in a closed cell whose temperature $T_0$ (100-300~K) can be regulated over the course of several weeks within a 10~mK range. Two modes of operation are used: in a measurement mode (see fig.~\ref{fig1}c) the electrodes form a capacitor with glycerol as the dielectric. By applying a sinusoidal voltage at frequency $f$, the evolution of the dielectric permittivity of the liquid ($\epsilon^{\prime}(f)$ and $\epsilon^{\prime\prime}(f)$) can be measured using a lock-in amplifier (SR830). In the heating mode (see fig.~\ref{fig1}d) the electrodes are connected and a current $I(t)$ can flow through them and generate a heat flux $j(t)\propto I^2(t)$ heating the liquid. This current is delivered by an external large capacitor previously charged at a voltage $U_\mathrm{h}$.

\begin{figure}[htbp]
 
\centering
 
\includegraphics[width=\columnwidth]{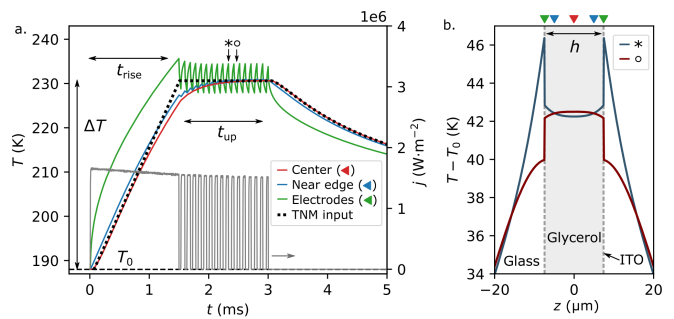}
 \caption{(a) Temperature evolution of the liquid during a heating-cooling sequence for $t_\mathrm{rise} = 1.5$~ms (at the electrode in green, near it in the glycerol in blue, and at half distance between the electrodes in red, see triangles in (b)), obtained by a numerical simulation from the heat flux dissipated at each electrode (shown in grey), deduced from the measured current $I(t)$. The dashed curve is the idealized temperature evolution used as input for the TNM model. (b) 1D temperature profile of the sample at two times denoted in (a) by symbols $\ast$ and $\circ$.}
  \label{fig2}
\end{figure}

\textbf{Temperature evolution and its estimation through thermal numerical simulation.} 
The temperature evolution of the liquid during a heating-cooling sequence could not be measured directly and was estimated using a 1D finite element numerical simulation (using FreeFem++~\cite{freefem}, see details in the supplementary) with thermal parameters for glass~\cite{Astrath_2005} and glycerol~\cite{PhysRevB.34.1631} taken from the literature and taking into account the thermal contact resistance between glycerol and electrodes~\cite{minakov2003advanced}. All the experiments shown in this work follow the same sequence: starting from thermal equilibrium at $T_0$, the liquid temperature is raised to $T_0+\Delta T$ (with $\Delta T = 34 - 45$~K) in a time $t_\mathrm{rise}$ (equal to 0.7, 1.5 or 6~ms) by applying an almost constant heat flux. This temperature is then maintained during a time $t_\mathrm{up}$ (from 0.7 to 10~ms) inside a range of a few kelvins by applying a previously determined (from a numerical simulation) sequence of heating pulses. The heating is then stopped and, the heated volume being very small, the liquid cools down quickly and reaches $T_0$ within 1~K after 1~s when the measurement process starts. Typical heat flux and temperature profiles are shown in fig.~\ref{fig2}a (for $t_\mathrm{rise} = 1.5$~ms, see supplementary for 0.7 and 6~ms). The temperature profile across the thickness of the liquid film is displayed in fig.~\ref{fig2}b at two instants of the constant high temperature phase. The temperature at the center of the sample stays nearly constant across the sample thickness due to the thermal contact resistance (within 0.2 K). The dashed black curve in fig.~\ref{fig2}a shows the idealized temperature profile used as input for the TNM model in section~\ref{section_TNM_model} for the sake of simplicity. In reality, the temperature elevation is a few kelvins lower than $\Delta T$ during the first millisecond. Besides, given the uncertainty on the simulation parameters the absolute determination of $\Delta T$ is subject to an uncertainty of less than 1.2~K (see supplementary). As discussed at the end of section~\ref{section_results}, the error introduced on the response time is too small to qualitatively affect the results of this work.
Experimentally, the parameters $t_\mathrm{rise}$ and $t_\mathrm{up}$ are controlled by the heat pulses sequence and the temperature elevation $\Delta T$ by the voltage $U_\mathrm{h}$.

\section{Results}
\label{section_results}
\begin{figure}[htbp]
\centering
\includegraphics[width=\columnwidth]{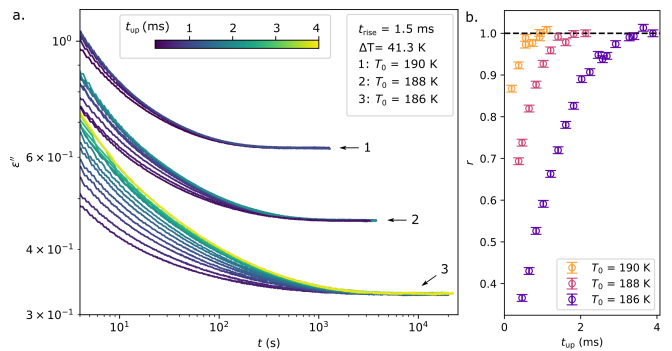}
 \caption{(a) Dielectric response $\epsilon^{\prime\prime}$ at $f=33$~Hz of the liquid after a heating-cooling sequence started at $t=0$ with $t_\mathrm{rise} = 1.5$~ms and $\Delta T = 42.6$~K for different values of $t_\mathrm{up}$ and for three equilibrium temperature $T_0$. (b) Ratio $r$ for the data shown on (a) as a function of $t_\mathrm{up}$.}
 
\label{fig3}
\end{figure}

Heating-cooling experiments were systematically performed on a glycerol sample starting from equilibrium states at temperatures $T_0 = 190$, $188$, and $186$~K. The evolution of the dielectric loss $\epsilon^{\prime\prime}(t)$ at $f = 33$~Hz was measured after the sample reached a thermal equilibrium (\textit{i.e.} $T-T_0 < 100$~mK for $t>10$~s). Data corresponding to $t_\mathrm{rise} = 1.5$~ms and $\Delta T = 42.6$~K are shown in fig.~\ref{fig3}a for different values of $t_\mathrm{up}$ and for the three equilibrium temperatures $T_0$. They display the classic behavior of an aging experiment: $\epsilon^{\prime\prime}(t)$ decreases even long after the sample has reached thermal equilibrium and eventually reaches the equilibrium value $\epsilon^{\prime\prime}_\mathrm{eq}(T_0)$. This process is faster for higher $T_0$ as the final equilibrium relaxation time $\tau_\alpha$ of the liquid gets shorter. An interesting feature that is specific to these heating-cooling experiments is the influence of $t_\mathrm{up}$: when this duration is increased from a small value, the aging curves $\epsilon^{\prime\prime}_{t_\mathrm{up}}(t)$, while always converging toward the same final value, seems to decrease from an out of range initial point of increasingly high value. For longer values of $t_\mathrm{up}$, however, all aging curves collapse on a maximum curve ($\epsilon^{\prime\prime}_{t_\mathrm{up}}(t) = \epsilon^{\prime\prime}_\mathrm{max}(t)$). On the range of $t$ observed, all the $\epsilon^{\prime\prime}(t) - \epsilon^{\prime\prime}_\mathrm{eq}$ curves (for a given value of $T_0$) appear identical to each other up to a multiplicative coefficient. This allows to characterize the effect of $t_\mathrm{up}$ using only one coefficient, denoted $r$ and defined as:
\begin{equation}
    r = \frac{\epsilon^{\prime\prime}_{t_\mathrm{up}}(t) - \epsilon^{\prime\prime}_\mathrm{eq}}{\epsilon^{\prime\prime}_\mathrm{max}(t) - \epsilon^{\prime\prime}_\mathrm{eq}}
\label{eq_def_r}
\end{equation}
For each set of parameters ($T_0$, $t_\mathrm{rise}$, $\Delta T$), a maximum curve $\epsilon^{\prime\prime}_\mathrm{max}(t)$, independent of $t_\mathrm{up}$, is determined and used to fit the ratio $r$ for all the other curves. The resulting ratios are shown in fig.~\ref{fig3}b as a function of $t_\mathrm{up}$ for the data shown in fig.~\ref{fig3}a and display in a clearer way the behavior qualitatively described above: $r$ increases and converges to 1. While this is observed for all values of $T_0$, $t_\mathrm{rise}$, $\Delta T$, the time $t_\mathrm{up}$ at which the plateau is reached strongly depends on these parameters.

\begin{figure}[htbp]
 
\centering
 
\includegraphics[width=\columnwidth]{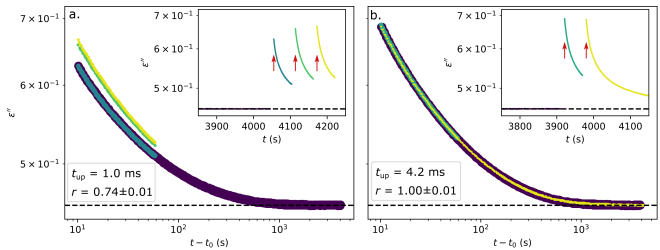}
 \caption{Dielectric response $\epsilon^{\prime\prime}(t-t_0)$ at $T_0=188$~K after heating-cooling sequences occurring at $t_0$ with $t_\mathrm{rise}=1.5$~ms, $\Delta T = 38.6$~K, and 2 values of $t_\mathrm{up}$ corresponding to $r=0.74$ (a) and $r\approx 1$ (b). The insets show the same response as a function of $t$ with red arrows denoting the heating-cooling sequences.}
 
\label{fig4}
\end{figure}

To get a better insight on this $t_\mathrm{up}$ dependence, a few experiments were performed starting from non-equilibrated initial states (although thermally equilibrated) for $T_0=188$~K, $t_\mathrm{rise} = 1.5$~ms, $\Delta T = 39.8$~K and two values of $t_\mathrm{up}$. Fig.~\ref{fig4}a corresponds to $t_\mathrm{up}$ short enough so that $r$ does not reach 1 ($r\approx 0.74$). The dark blue curve is the reference obtained from an equilibrated initial state. Then, after a time $t = 4000$~s long enough for the system to be equilibrated ($\tau_\alpha(188~K) \approx 100$~s), the same heating-cooling sequence is applied, leading to the dark green curve that falls exactly on the reference curve which illustrates the reproducibility of these experiments. Next, after $\approx 70$~s and while the system has not yet reached equilibrium ($\epsilon^{\prime\prime}(t)>\epsilon^{\prime\prime}_\mathrm{eq}$), a new sequence is initiated from which the green curve is obtained, and, finally, this process is repeated after $70$~s leading to the yellow curve. These two aging curves obtained from non-equilibrated initial states are both different and do not collapse on the reference curve (see fig.~\ref{fig4}). This shows that after the heating-cooling sequence, the system has kept a memory of its initial state. Fig.~\ref{fig4}b displays a similar set of experiments for which $t_\mathrm{up}$ is now long enough so that the aging curve has reached $\epsilon^{\prime\prime}_\mathrm{max}(t)$ (so that $r = 1$). In this case, the aging curves are the same whether or not the initial state is equilibrated meaning that the system does not appear to keep a memory of its initial state.

These results allow us, without relying on a model yet, to come up with the following interpretation of the liquid response to a heating-cooling experiment: from an initially equilibrated state at $T_0$, the system is perturbed and set out of equilibrium by a sudden rise in temperature ($t_\mathrm{rise} \ll \tau_\alpha(T_0)$). Then the system is maintained during $t_\mathrm{up}$ at high temperature at which $\tau_\alpha(T_0+\Delta T)$ is at least a factor 10 smaller than $t_\mathrm{up}$. This time can be long enough for the liquid to fully equilibrate at $T_0+\Delta T$, in which case the aging curve $\epsilon^{\prime\prime}_\mathrm{max}(t)$ becomes, as observed, independent of $t_\mathrm{up}$ and does not keep the memory of the initial state. However, due to the very high temperature step ($> 30$~K) and the non-linear response of supercooled liquids, $t_\mathrm{up}$ can be too short to allow a full equilibration during the high temperature phase in which case $\epsilon^{\prime\prime}(t)$ depends on the non-equilibrated state reached at the end of this phase which itself is affected by the initial state (if not at equilibrium).

\begin{figure}[htbp]
\centering
\includegraphics[width=\columnwidth]{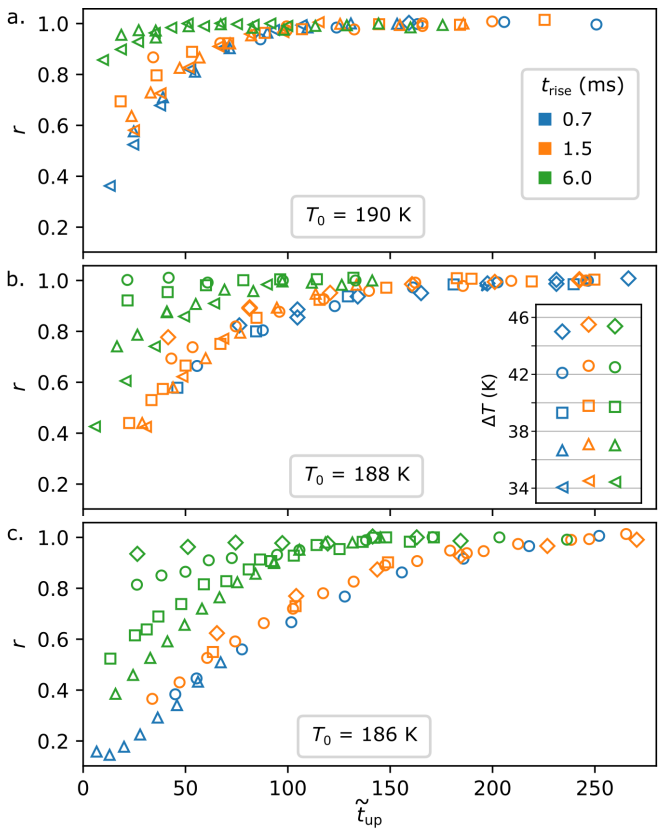}
 \caption{Ratio $r$ as a function of the the dimensionless time $\Tilde{t}_\mathrm{up}$ spent at temperature $T_0+\Delta T$. All the data gathered in this study are shown here. The three equilibrium temperatures $T_0$ correspond to the subfigures a, b, and c. The three values of the heating phase duration $t_\mathrm{rise}$ are shown with colors. The amplitude of the temperature steps $\Delta T$ is distinguished with marker shapes with the legend displayed in the inset of (b). The relative uncertainty is of the order of the size of the markers.}
\label{fig5}
\end{figure}

The very different $t_\mathrm{up}$ needed to reach $r=1$ for various $T_0$ visible in fig.~\ref{fig3}b results from the fact that the temperatures $T_0+\Delta T$ at which this phenomenon occurs are associated with very different relaxation rate. In the following, we chose, in order to compare more easily the different datasets, to consider a dimensionless duration of the high temperature phase, $\Tilde{t}_\mathrm{up}$, defined as:
\begin{align}
    \Tilde{t}_\mathrm{up} = \frac{t_\mathrm{up}}{\tau_\alpha(T_0+\Delta T)}
\end{align}
where $\tau_\alpha(T)$ is the dielectric $\alpha$ time (see fig. S2c of the suppl. mat.). Physically, this corresponds to measuring the time spent at $T_0+\Delta T$ with the material time that the liquid would have if it was equilibrated from the start at this temperature (although it is not). This re-scaling is used in fig.~\ref{fig5} to plot the ratio $r$ for all the heating-cooling experiments as a function of $\Tilde{t}_\mathrm{up}$. All the curves follow the same previously described trend with $r$ converging to 1 at long times. The dimensionless time at which the liquid equilibrates at $T_0+\Delta T$ is high (in the range $100-200$) which means that the time needed for the liquid to equilibrate at this temperature is much larger than the equilibrium relaxation time. This time depends on the parameters of the sequence: it is longer when the rising time $t_\mathrm{rise}$ is shorter and when the equilibrium temperature $T_0$ is lower. For $t_\mathrm{rise} = 6$~ms, the amplitude of the temperature step $\Delta T$ affects the way the liquid approach equilibrium with higher steps leading to higher ratio values at short times. On the contrary for $t_\mathrm{rise} = 0.7$ and 1.5~ms, the behavior appears less dependent on $\Delta T$. The uncertainty associated with the measurement of the equilibration time can be quantified: the use of an idealized temperature profile leads to an error $<5$~\% for $t_\mathrm{rise} = 1.5$~ms and $<10$~\% for $t_\mathrm{rise} = 0.7$~ms (see supplementary), while the uncertainty on the absolute determination of $\Delta T$ corresponds to a $\pm 30$~\% error on the relaxation time. The lowest estimation of the dimensionless time of the liquid equilibration at $T_0+\Delta T$ thus remains very large compared to 1.

\section{Application of the TNM model}
\label{section_TNM_model}
\subsection{Fictive temperature}
A useful way of quantifying, in temperature unit, the distance from equilibrium of a supercooled liquid is the fictive temperature, denoted as $T_\mathrm{f}$ which expresses the structural contribution (by subtracting the glassy contribution) in the evolution of a macroscopic quantity (such as the density, the enthalpy, or here the dielectric loss $\log \epsilon^{\prime\prime}$)~\cite{mauro2009fictive}. As the measurements are performed only at the end of the cooling regime, the variation of this quantity is small enough to be linearized: $\Delta \log \epsilon^{\prime\prime} = \alpha_\mathrm{gl}\Delta T + \alpha_\mathrm{s}\Delta T_\mathrm{f}$ where $\alpha_\mathrm{s}$ expresses the effect of structural rearrangements on $\log \epsilon^{\prime\prime}$ while $\alpha_\mathrm{gl}$ corresponds to the glassy (\textit{i.e.} isotructural) part. We measured these two coefficients (see supplementary) and plot the evolution of $T_\mathrm{f}$ at the end of the cooling phase in fig.~\ref{fig6}b (red markers). This conversion makes the comparison with the prediction of the TNM model much easier.

\subsection{Parameters of the TNM model}
The application of the TNM model requires the knowledge of two functions that characterize the liquid's response: the memory kernel $M(\xi)$ and the out-of-equilibrium relaxation rate $\gamma(T, T_\mathrm{f})$. A stretched exponential function~\cite{mazurin1977relaxation} can be used as a kernel: 
\begin{equation}
    M(\xi) = e^{-(\xi/\xi_0)^\beta}, \quad \xi_0 = \frac{1}{\int_0^\infty e^{-u^\beta}du}
    \label{eq_M}
\end{equation}
with as adjustable parameter the exponent $\beta$. The coefficient $\xi_0$ (close to 1) is present in order to ensure that the memory kernel corresponds to a unity relaxation time ($\int_0^\infty M(\xi)d\xi = 1$) independently of the value of $\beta$. It is not classic in the literature, but it is for example needed if one wants to adjust with a stretched exponential the memory kernel that can be deduced from the measurements of Roed~\textit{et al.}~\cite{roed2019generalized} on glycerol (see supplementary) or of Riechers~\textit{et al.}~\cite{riechers_predicting_2022} on another molecular glassformer.

For the relaxation function $\gamma(T, T_\mathrm{f})$, the phenomenological forms introduced first by Narayanaswamy~\cite{narayanaswamy1971model} was :
\begin{align}
\gamma(T, T_\mathrm{f}) = \frac{1}{\tau_\mathrm{eq}^{x}(T) \tau_\mathrm{eq}^{1-x}(T_\mathrm{f})}
    \label{eq_gamma_T_Tf}
\end{align}
with $\tau_\mathrm{eq}(T)$ an arrhenian relaxation rate ($\log \tau_\mathrm{eq}(T) \propto 1/T$) and $x\in[0,1]$ a nonlinearity parameter~\cite{moynihan1976structural}. While this form was successfully used in several studies~\cite{moynihan1976structural, hodge1994enthalpy, mauro2009nonequilibrium}, the arrhenian expression of $\tau_\mathrm{eq}(T)$ makes it not suited for fragile glass formers (such as glycerol) that display a dramatic increase in equilibrium relaxation time when decreasing $T$ close to $T_\mathrm{g} = 187$~K. To remedy this, we choose to use instead a VFT form for the relaxation time:
\begin{align}
\tau_\mathrm{eq}(T) = \tau_0 \exp\left(\frac{A}{T-T_\mathrm{K}}\right)
\label{eq_VFT}
\end{align}
with parameters adjusted on dielectric data of $\alpha$ relaxation from ref.~\cite{lunkenheimer_dielectric_2002} ($\tau_0=6.5\cdot 10^{-15}$~s, $A=2420$~K, $T_\mathrm{K} = 127$~K, as shown in the supplementary, our measured values of $\tau_\alpha$ agree very well with this VFT law). The expression that we used in this work for $\gamma(T, T_\mathrm{f})$ is thus the combination of eq.~\ref{eq_gamma_T_Tf} and eq.~\ref{eq_VFT}. The physical meaning of eq.~\ref{eq_gamma_T_Tf} and of the nonlinearity parameter $x$ can be expressed in the following way: out of equilibrium, the mean relaxation rate is affected both by the structure (\textit{i.e.} depends on $T_\mathrm{f}-T$: the more out of equilibrium the system is, the faster it relaxes) and by the temperature of the phonons $T$ (which changes the density: a very out of equilibrium structure relaxes slower if it is denser). The relative importance of these two effects is expressed by $x$ (with the limiting cases being $\gamma|_{x=0} = \gamma(T_\mathrm{f})$ and $\gamma|_{x=1} = \gamma(T)$). The fact that we use an expression of $\tau_\alpha(T)$ valid on a very wide range of temperatures, rather than a local linearization of $\log \tau_\alpha(T)$ (as in ref.~\cite{hecksher2015communication,roed2019generalized}) is required given the large amplitude of the temperature steps that we apply ($T_\mathrm{f}-T \in [-45~\mathrm{K}, +6~\mathrm{K}]$). For this purpose, other phenomenological forms are available in the literature~\cite{mazurin1983temperature, mauro2009nonequilibrium} and the one chosen by Mazinani~\textit{et al.}~\cite{mazinani2012enthalpy}, while not strictly equivalent, would lead to very similar results. For small enough $T_\mathrm{f}-T$, these forms are usually all equivalent, in which case a simpler first-order Taylor expansion of $\log \gamma(T, T_\mathrm{f})$ can be used~\cite{hecksher2015communication, roed2019generalized, riechers_predicting_2022} with the advantage of necessitating no hypothesis on the $T_\mathrm{f}$ dependence of $\gamma$. This is the option chosen by Roed~\textit{et al.}~\cite{roed2019generalized} for their ideal step experiments on glycerol, with amplitudes between 2 and 8~K. From their results, a nonlinearity parameter $x_\mathrm{S} = 0.18 \pm 0.09$ can be deduced (see details in supplementary). For the reasons mentioned above, this value can be used with our expression of $\gamma$ for small distances from equilibrium.

\subsection{Application of the model}
In this work, the input of the model is an idealized version of the temperature evolution $T(t)$ of the liquid, determined by the thermal numerical simulation, shown as a dashed black line in fig.~\ref{fig2}a for $t_\mathrm{rise} = 1.5$~ms (see supplementary for the other values of $t_\mathrm{rise}$) in which the temperature elevation reaches $\Delta T$ at $t=t_\mathrm{rise}$ and is then maintained constant for a duration $t_\mathrm{up}$. In the TNM model, the linear response of eq.~\ref{eq_TNM_resp} is expressed in the material time domain and the variation of $T(t)$ is here introduced by computing $\frac{\mathrm{d}T}{\mathrm{d}\xi} = \frac{\mathrm{d}T}{\mathrm{d}t}/\gamma(T, T_\mathrm{f})$ at each time step.

\begin{figure}[htbp]
\centering
\includegraphics[width=\columnwidth]{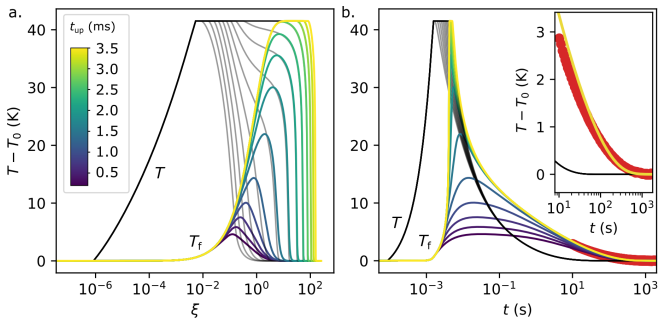}
 \caption{Application of the TNM model (for $T_0 = 188$~K, $t_\mathrm{rise} = 1.5$~ms and $\Delta T = 41$~K), with parameters $\beta=0.8$, $x_{\uparrow} = x_\mathrm{L} = 0.465$ and $x_{\downarrow} = x_\mathrm{S} = 0.18$, showing the system temperature $T$ (black) and fictive temperature $T_\mathrm{f}$ (color) as a function of the material time (a) and the observer time (b). Experimental measurements are shown in red. The inset is a zoom on the experimentally accessible time range.}
\label{fig6}
\end{figure}

The model is applied for the heating-cooling sequence for different values of $t_\mathrm{up}$. Typical results of the input $T$ and output $T_\mathrm{f}$ are shown in fig.~\ref{fig6} as a function of the material time $\xi$ (a) and of the observer time $t$ (b). It is already visible that the TNM model allows to reproduce the observed features of fig.~\ref{fig3}a: long enough values of $t_\mathrm{up}$ lead to a collapse of the aging curves at the end of the cooling phase while they fall below for shorter values of $t_\mathrm{up}$. The evolution of $T_\mathrm{f}$ confirms the interpretation of the previous section: the aging curves measured for $t>10$~s contain quantitative information on what happens during the first few milliseconds of the heating phase and the observed influence of $t_\mathrm{up}$ is related to the degree of equilibration during the high temperature phase. For $t>10$~s, similarly to the experimental data, the curves appear almost proportional to each other which allows to compute the ratio $r$ by following the same procedure described earlier for the experimental data. The comparison of the experimental and modeled $r(\Tilde{t}_\mathrm{up})$ curves can be used to adjust the parameters of the model. However, it appears that it is not possible to reproduce quantitatively every aspect of the observations with only one value of the nonlinearity parameter $x$. Indeed, the value measured in glycerol by Roed~\textit{et al.}~\cite{roed2019generalized} $x_\mathrm{S} = 0.18$, leads to a fairly good agreement with the experimental data for the long-term relaxation $\epsilon^{\prime \prime}(t > 10~\mathrm{s})$ but, predicts an equilibration at $T_0+\Delta T$ after $\tilde{t}_\mathrm{up} > 10^4$. This is the sign of an overestimation of the weight of $T_\mathrm{f}$ compared to $T$ on $\gamma$, \textit{i.e.} $x$ being too small. Conversely, $x_\mathrm{L} = 0.465$ predicts the right $r(\tilde{t}_\mathrm{up})$ dependence but leads to a slower relaxation of $\epsilon^{\prime \prime}(t > 10~\mathrm{s})$ than what is measured (see fig.~S6b of the supplementary). This can be interpreted as an underestimation of the weight of $T_\mathrm{f}$, $x$ being too large.

In order to get a quantitative agreement with the experiments while keeping the model as simple as possible, we chose to use two values of the nonlinearity parameter: $x_{\uparrow}$ during the heating phase ($T>T_\mathrm{f}$) and $x_{\downarrow}$ during the cooling phase ($T<T_\mathrm{f}$). This makes $\gamma(T, T_\mathrm{f})$ non-differentiable on the $T=T_\mathrm{f}$ line but does not induce a discontinuity. Considering the asymmetry of our experiment ($T_\mathrm{f} -T $ goes as down as -45~K in the heating phase but stays below 6~K during cooling), this can be viewed as a simple way to apply different values of $x$ depending on the distance for equilibrium. We chose to set the value of $x_{\downarrow}$ to $x_\mathrm{S} = 0.18$ measured by Roed~\textit{et al.}~\cite{roed2019generalized} for distance to equilibrium $<8$~K and we used $x_{\uparrow} = x_\mathrm{L}$, corresponding to large distances from equilibrium, as a free parameter. Each fitting parameter ($\beta$ and $x_{\uparrow}$) has a separable effect on the model prediction for $r(\tilde{t}_\mathrm{up})$ (see fig.~S8 of the supplementary): changing $x_{\uparrow}$, even by a small amount, shifted the curve along the horizontal axis while $\beta$ had a shift effect and an influence on the smoothness of the curve. The best overall agreement was obtained with $\beta = 0.8$ and  $x_{\uparrow} = 0.465$. The inset of fig.~\ref{fig6}b shows a decent agreement on the cooling phase (see also fig.~S6 of the supplementary) given the fact that the value of $\beta$ was chosen to adjust the response to heating rather than this part of the data. Fig.~\ref{fig7}a displays the result of the model with these parameters for $T_0 = 188$~K, all three values of $t_\mathrm{rise}$ and several values of $\Delta T$. A way to quantitatively compare the model and experimental data is to consider the value $\Tilde{t}_{\mathrm{up}|r=0.9}$ needed for $r$ to reach 0.9 (dashed horizontal lines on fig.~\ref{fig7}a). The diagrams of fig.~\ref{fig7}b display this threshold plotted as a function of $\Delta T$ while experimental data extracted from fig.~\ref{fig5} are shown with markers. While the agreement between the two is not perfect, two features are well reproduced: first, the fact that $\Tilde{t}_{\mathrm{up}|r=0.9}$ increases when $t_\mathrm{rise}$ is decreased, but also its non-monotonic behavior with $\Delta T$. For $t_\mathrm{rise} = 1.5$~ms, the experimentally observed maximum near $\Delta T \approx 40$~K is correctly predicted as well as the increases with $\Delta T$ for $t_\mathrm{rise} = 0.7$~ms and the decreases for $t_\mathrm{rise} = 6.0$~ms. The model with the same parameters was also applied for other values of $T_0$ with results shown in fig.~\ref{fig7}c. If the effect of the decrease of $\Tilde{t}_{\mathrm{up}|r=0.9}$ with $T_0$ is qualitatively reproduced, the model fails to capture the correct dependence with $T_0$ and overestimates its effect. 

The value of $\beta = 0.8$ is significantly higher than the 0.55 value that can be deduced from Roed~\textit{et al.}~\cite{roed2019generalized} measurements performed in the 176-184~K range. This could be explained by the fact that the memory kernel should not be a constant function on a wide range of temperatures. Indeed, the dielectric spectra measured at equilibrium on a wide range of temperatures by Lunkenheimer and Loidl~\cite{lunkenheimer_dielectric_2002} shows an increase in the slope $\beta_\mathrm{CD}$ of the high-frequency wings from 0.55 at 180~K to 0.60 at 250~K and 0.75 at 400~K. The stretched exponent coefficient can be linked to this quantity by $\beta \approx (\beta_\mathrm{CD})^{0.8}$~\cite{alvarez1991colmenero} but this is still not enough to explain our high value of $\beta$. In 1991, Moynihan~\textit{et al.}~\cite{moynihan1991linear}, already noticed that the application of the TNM model to calorimetry experiments on glycerol required, for high heating rate corresponding to more out-of-equilibrium conditions, the use of higher values of $\beta$ that the one measured at equilibrium on the same temperature range by linear experiments. This illustrates the limits of the TNM model when considering far-from-equilibrium states.

The need for the two values of the nonlinearity parameters can be interpreted in the following way: for very small distances from equilibrium, the role of $T$ and $T_\mathrm{f}$ on $\gamma(T, T_\mathrm{f})$ is not distinguishable and the response is linear ($x$ plays no role). For larger distances, the relaxation rate depends on both these temperatures and it make sense to try to simply describe this dependence, as the TNM model does, with only one parameter $x$. This assumption that $x$ is independent of the sign or the amplitude of $T-T_\mathrm{f}$ is expected to be valid only in a small enough range of $T-T_\mathrm{f}$. For larger distances from equilibrium, the relative weight of $T$ and $T_\mathrm{f}$ in $\gamma(T, T_\mathrm{f})$ could depend on the distance and its sign. Our findings are consistent with the observation that deviations from the TNM model (with a single parameter $x$) start to be perceptible for ideal temperature steps of 8~K for glycerol~\cite{roed2019generalized}, and 3~K on another molecular glass former~\cite{riechers_predicting_2022}. The form of $\gamma(T, T_\mathrm{f})$ is mentioned as one of the possible interpretations of these deviations~\cite{riechers_predicting_2022}. Our approach here, given our large but limited range of $\Delta T$, is to use one effective value of $x_{\uparrow} = x_\mathrm{L}$ that contains a mean information on the form of $\gamma(T, T_\mathrm{f})$ for this typical distance from equilibrium. The observation that $x_\mathrm{L} \neq x_\mathrm{S}$ supports the above statement while the deviation between the experimental data and the model concerning the effect of $\Delta T$ could be seen as a consequence of the simplification of using a mean value of $x_\mathrm{L}$ for the whole range of $\Delta T = 34 - 46$~K. Finally, it is interesting to note that the values of the nonlinearity parameter that we used ($x_\mathrm{S} = 0.18$ and $x_\mathrm{L} = 0.465$) fall around the RFOT prediction~\cite{lubchenko2004theory} ($x\approx 19/m = 0.35$, with $m = 53$ the fragility of glycerol).

The fact that the model fails to account for the right dependence on $T_0$ can also be understood in this context. While it is \textit{a priori} surprising given the fairly small $T_0$ range studied (186-190~K, although it still corresponds to a decade of change in relaxation time), it has to be noted that the model prediction is highly sensitive to the value of $x_{\uparrow}$. Indeed, a better agreement would be obtained at $T_0 = 190$~K with $x_{\uparrow} = 0.44$ and at $T_0 = 186$~K with $x_{\uparrow} = 0.49$ which corresponds to small changes (compared to the difference with $x_\mathrm{S}$). This could be interpreted as the fact that for large distances from equilibrium, $x$ should not only depend on $\Delta T$ but also on the absolute value of the fictive temperature $T_\mathrm{f}(t=0) = T_0$.

\section{Discussion}
\label{section_discussion}
\begin{figure}[htbp]
\centering
\includegraphics[width=\columnwidth]{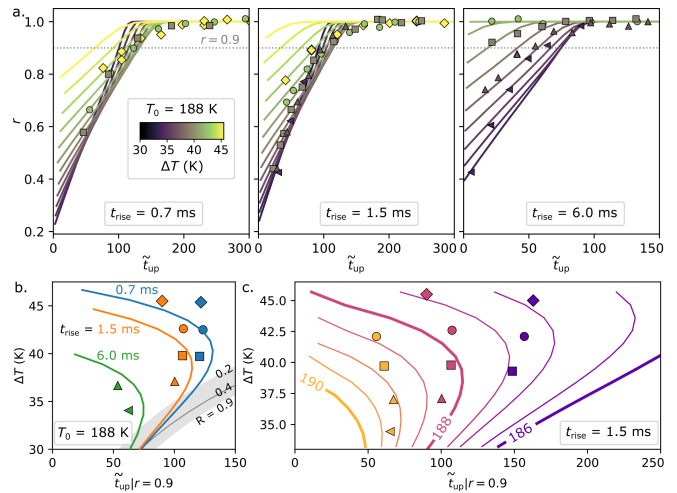}
 \caption{Comparison of experimental data and the TNM model (with parameters $\beta = 0.8$, $x_{\uparrow} = x_\mathrm{L} = 0.465$ and $x_{\downarrow} = x_\mathrm{S} = 0.18$). (a) Same plot as fig.~\ref{fig5}b for $T_0= 188$~K with the model as solid lines and experimental data as markers. The colors denote the heating amplitude $\Delta T$. (b, c) Diagram of the dimensionless duration $\tilde{t}_\mathrm{up}|_{r=0.9}$ needed to reach $r=0.9$ (dashed lines on (a)) as a function of $\Delta T$.  (b) At $T_0 = 188$~K and for 3 values of $t_\mathrm{rise}$. The grey zone corresponds to ideal step experiments and constant values of $R$ (defined in eq.~\ref{eq_def_R}) rather than $r$ (top border: $R=0.2$, darker grey line: $R=0.4$, bottom border: $R=0.9$). (c) For $t_\mathrm{rise} = 1.5$~ms and $T_0$ in the range 186-190 K with 0.5~K increment. The experimental data are shown with markers and the model with solid lines.
 }
\label{fig7}
\end{figure}

Despite the limitations discussed above, it is remarkable that the TNM model, with only 2 adjustable parameters, allows us to describe with such a decent agreement the experimental observation on the response of a liquid up to 45~K temperature steps. As the model gives access to the whole response $T_\mathrm{f}(t)$, it can be used to better understand our experimental observations.

\begin{figure}[htbp]
\centering
\includegraphics[width=\columnwidth]{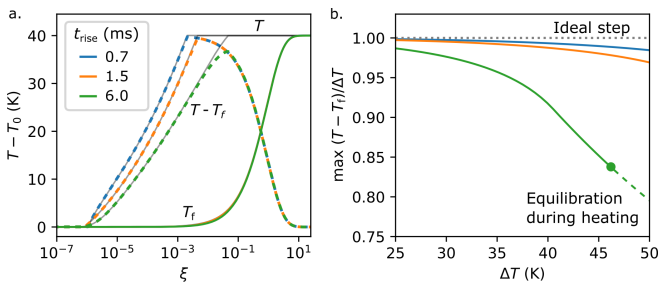}
 \caption{Effect of $t_\mathrm{rise}$ on the upward temperature step computed using the TNM model (with $\beta = 0.8$ and $x_{\uparrow} = x_\mathrm{L} = 0.465$) at $T_0 = 188$~K. (a) Response $T_\mathrm{f} -T_0$ (color lines) of the liquid to temperature steps $T-T_0$ (grey lines) of amplitude $\Delta T = 40$~K as a function of the material time. The difference $T-T_\mathrm{f}$ is shown with dashed lines. (b) Maximum amplitude of $T-T_\mathrm{f}$ as a function of $\Delta T$ (solid lines). The dashed line corresponds to the case in which the liquid equilibrates during the heating phase. The dotted grey line corresponds to an ideal step experiment.}
\label{fig8}
\end{figure}

One of them is the effect of $t_\mathrm{rise}$: smaller values lead to higher $t_\mathrm{up}$ duration needed for the system to equilibrate. A small $t_\mathrm{rise}$ compared to $\tau_\alpha(T_0+\Delta T)$ would be a sufficient (although not necessary) condition to realize an ideal step experiment such as the one performed by Hecksher~\textit{et al.}~\cite{hecksher2010physical, hecksher2015communication}. Conversely, an extremely long value would lead to the equilibration of the liquid during the heating phase, before the maximum temperature is reached. In this work, we are in an intermediate regime as $\tau_\alpha(T_0 + \Delta T) < t_\mathrm{rise} \ll \tau_\alpha(T_0)$ in which it is hard to characterize \textit{a priori} the ideality of the step experiments. For this, the TNM model can be helpful. By studying the model prediction, shown in fig.~\ref{fig8}a, during the heating and constant temperature phases (with the parameters determined above) for $T_0 = 188$~K, $\Delta T = 40$~K and three values of $t_\mathrm{rise}$, it appears that the distance from equilibrium $T - T_\mathrm{f}$ (shown as dashed lines) reaches 99.0~\% of $\Delta T$ for $t_\mathrm{rise} = 0.7$~ms at the end of the heating phase. This is thus very close to an ideal step experiment for which at $t=0$, $T - T_\mathrm{f} = \Delta T$ by definition. For $t_\mathrm{rise} = 6.0$~ms, however, $T-T_\mathrm{f}$ reaches at maximum only 90~\% of $\Delta T$, meaning that the liquid had started to relax during the heating phase. The ratio $\max (T - T_\mathrm{f})/\Delta T$ is displayed in fig~\ref{fig8}b as a function of $\Delta T$. Our experiments with $t_\mathrm{rise} = 0.7$ and 1.5~ms are very close to ideal step experiments (with $<1$~\% relaxation the whole range of $\Delta T$ for the former and for $\Delta T<39$~K for the later). Experiments for which $t_\mathrm{rise} = 6$~ms were not only all far from being ideal in the range studied but even almost reached the limit under which they could not be called step experiments as for $\Delta T>45$~K the liquid equilibrates before the end of the heating phase. Note that the quantity plotted is the maximum distance from equilibrium reached during heating and not the final $t=t_\mathrm{rise}$ value, which is almost null in this last case. All this allows to explain the experimental observations of fig.~\ref{fig5} as consequences of the non-ideality of the step experiments. Indeed, the observer time $t$, and thus $\tilde{t}_\mathrm{up}|_{r=0.9}$, results from the integration of $1/\gamma(T, T_\mathrm{f})$. For an ideal step, this quantity starts at $1/\gamma(T_0+\Delta T, T_0)$  (at the end of the instantaneous heating phase) from which it decreases to $1/\gamma(T_0+\Delta T, T_0+\Delta T)$ when equilibrium is reached. For non-ideal steps, however, the starting value will be smaller, leading to shorter $\tilde{t}_\mathrm{up}|_{r=0.9}$. 

Another experimental observation is the non monotonic effect of $\Delta T$ on $\Tilde{t}_\mathrm{up}|_{r=0.9}$ visible in fig.~\ref{fig7}b and c. It is important to emphasize that we do not have access experimentally to $T_\mathrm{f}(t_\mathrm{up})$ but instead, we use $r(t_\mathrm{up})$ as a proxy to access it. We arbitrarily chose the threshold $r=0.9$ to define a time $\tilde{t}_\mathrm{up}|_{r=0.9}$ at which the equilibrium is almost reached at high temperature but the correspondence with $T_\mathrm{f}$ is not straightforward. From the application of the model shown in fig.~\ref{fig7} it is possible to link any value of $r$ with the degree of relaxation of the system at the end of the high temperature phase: 
\begin{align}
R = \frac{T_\mathrm{f}(t_\mathrm{up})-T_0}{\Delta T}
\label{eq_def_R}
\end{align}
Yet, the link between $r$ and $R$ appears strongly dependent on $\Delta T$: for $T_0 = 188$~K and $t_\mathrm{rise} = 0.7$~ms the threshold $r=0.9$ corresponds to $R = 0.25$ for 34.5~K but 0.03 for 45.5~K. This small value compared to the corresponding $0.9$ limit on $r$ is mainly due to the fact that the system is not instantly quenched after $t_\mathrm{up}$ from $T=T_0+\Delta T$ to $T_0$ but instead cooled at a rate that still allows some relaxation. This is amplified for greater temperature steps $\Delta T$ for which smaller relaxation times are reached. In fig.~\ref{fig7}b, the case of an ideal step experiment is shown with a grey area whose borders correspond to constant $R$ values of respectively 0.2 (top) and 0.9 (bottom) while the grey line inside corresponds to $R=0.4$. If $r=0.9$ corresponds to a given limit in $R \in [0.4, 1]$, the distinction between the two is of little importance (compared to the experimental uncertainty) and $r$ is a straightforward proxy to $R$. In this case, corresponding to relatively low $\Delta T$ (the upper limit  $30-35$~K is barely accessible here), the increasing $\Tilde{t}_\mathrm{up}|_{r=0.9}(\Delta T)$ directly reflects the temperature fictive effect. At higher $\Delta T$ (corresponding to the range studied) an opposing, and quickly dominating effect takes place as the $r=0.9$ limit corresponds to smaller and smaller $R$ limits.

\begin{figure}[htbp]
 
\centering
 
\includegraphics[width=\columnwidth]{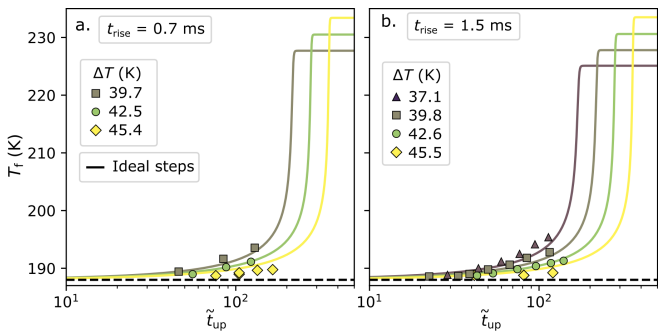}
 \caption{Reconstruction, from the experimental data (markers), limited to $r<0.96$, of the $T_\mathrm{f}$ evolution after an upward temperature step of amplitude $\Delta T$ starting at $T_0 = 188$~K, using the TNM model (with parameters $\beta = 0.8$, $x_{\uparrow} = x_\mathrm{L} = 0.465$ and $x_{\downarrow} = x_\mathrm{S} = 0.18$). The solid lines are the application of the TNM model for ideal steps with the same parameters. %
 }
\label{fig9}
\end{figure}

From this understanding, it is possible to go further and use the TNM model to determine the relation $R(r)$ for each experimental condition (see fig.~S9 of supplementary) and finally get access to the evolution of the fictive temperature during the upward step. The result is shown in fig.~\ref{fig9} as a function of $\Tilde{t}_\mathrm{up}$ with markers and is compared with the simple application $T_\mathrm{f}(\xi) = T_0 + \Delta T [1- M(\xi)]$ of the TNM model in the case of ideal temperature steps of same amplitude (solid lines). It is important to note that this reconstruction of $T_\mathrm{f}$ from the experimental data depends on the parameters of the model that were determined from these same data. Still, the self-consistency displayed by the fairly good agreement between the data and the simple version and the model shows that the TNM model, captures qualitatively the observations while showing its limit in predicting accurately the effect of $\Delta T$. Indeed, on the one hand,  this figure allows us to get a better understanding of the fact that only low values of $R$ are accessible by our experiment: the liquid response to such huge temperature steps displays a highly non-linear behavior as it is strongly affected by the fictive temperature effect. It can be broken into two parts: a long, and extremely $\Delta T$ dependent, phase of "awakening" needed for $T_\mathrm{f}$ to rise from $T_0$ by a few kelvins followed by an equilibration phase happening only in a few units of $\tau_\alpha(T_0+\Delta T)$, too fast to be accurately measured by our method. On the other hand, it is visible that the model, while giving an accurate description of the response for $\Delta T \approx 42.5$~K, is not as self-consistent for higher or smaller $\Delta T$. This again illustrates the limitation of using a $\Delta T$ independent nonlinearity parameter $x_{\uparrow}$ for such far-from-equilibrium states.

\section{Conclusion and perspectives.}
We presented in this work a quantitative experimental study of the response of a supercooled liquid to upward temperature steps with amplitude ranging from 34 to 45~K. These observations were made possible by the development of a new experimental device allowing us to heat a thin liquid layer within a millisecond time scale, at a rate between 5~000 and 60~000~K/s, and to control precisely the duration of the high temperature phase before cooling it down to the initial temperature. We showed that, by monitoring the final relaxation of the dielectric loss, it was possible to obtain quantitative information on the liquid response to the heating step. The observations were overall well captured by the TNM formalism despite the significant distance from equilibrium reached provided that different values for the nonlinearity parameter were used for the heating and cooling phase that corresponds to very different distances from equilibrium. This work challenges the TNM model by illustrating its limits, especially concerning the fact that the $T$ and $T_\mathrm{f}$ dependence of the relaxation rate can be described by only one parameter $x$, but also its strengths: it was able to provide a precise understanding on how the measurements performed in a 10 - 10 000~s timescale are related to the liquid response during the first milliseconds of the experiment and allowed to reconstruct this response in a self-consistent manner. 

Our work provides a phenomenological understanding of the duration needed for a supercooled liquid to equilibrate after a high amplitude temperature step and could be useful in designing ways to generate very unstable glasses, \textit{i.e.} quenched in a glassy state with high rate and corresponding to high fictive temperature. Such systems could display interesting mechanical properties~\cite{ozawa2018random} and help the search for Gardner transition in structural glasses~\bibnote[Note]{One of the arguments put forward by ref~\cite{albert2021searching} to explain the absence of a Gardner transition in glycerol is the generally well-annealed character of structural glasses (compared for example to colloids or granular systems). It would thus be interesting to search for Gardner transition in a poorly annealed (\textit{i. e.} unstable) glass.},\cite{albert2021searching}.

The TNM approach is based on the idea of a uniform fictive temperature and does not take into account any spatial heterogeneities in the relaxation. Yet, the fact that it described reasonably well our data does not allow us to decide between different relaxation mechanisms and to infirm or confirm, on glycerol, recent observations of a transition between cooperative relaxation for small step amplitude and growth of high-mobility regions for far-from-equilibrium states~\cite{vila2023emergence}. Such observations, in the range of small relaxation time explored here (as low as 10~$\upmu$s), and starting from equilibrated states, would be useful to get a better understanding of the nature of this transition. Adapting our experiment, to distinguish between these relaxation mechanisms will be the subject of future work.

\section*{Supplementary Material}
See supplementary material for details on the experimental setup, the thermal numerical simulations and the application of the TNM model.

\section*{Author's contribution}
MH and FL conceived the study. MH designed and performed the experiments and analyzed the data. MH wrote the manuscript in consultation with FL. Both authors contributed to the data interpretation and discussion of the results.

\begin{acknowledgments}
The authors are grateful to LABEX PALM and IRAMIS Institute for ﬁnancial support.
\end{acknowledgments}

\section*{Data Availability Statement}
The data that support the findings of this study are available from the corresponding author upon reasonable request.

%\bibliography{biblio}

\begin{thebibliography}{43}%
\makeatletter
\providecommand \@ifxundefined [1]{%
 \@ifx{#1\undefined}
}%
\providecommand \@ifnum [1]{%
 \ifnum #1\expandafter \@firstoftwo
 \else \expandafter \@secondoftwo
 \fi
}%
\providecommand \@ifx [1]{%
 \ifx #1\expandafter \@firstoftwo
 \else \expandafter \@secondoftwo
 \fi
}%
\providecommand \natexlab [1]{#1}%
\providecommand \enquote  [1]{``#1''}%
\providecommand \bibnamefont  [1]{#1}%
\providecommand \bibfnamefont [1]{#1}%
\providecommand \citenamefont [1]{#1}%
\providecommand \href@noop [0]{\@secondoftwo}%
\providecommand \href [0]{\begingroup \@sanitize@url \@href}%
\providecommand \@href[1]{\@@startlink{#1}\@@href}%
\providecommand \@@href[1]{\endgroup#1\@@endlink}%
\providecommand \@sanitize@url [0]{\catcode `\\12\catcode `\$12\catcode
  `\&12\catcode `\#12\catcode `\^12\catcode `\_12\catcode `\%12\relax}%
\providecommand \@@startlink[1]{}%
\providecommand \@@endlink[0]{}%
\providecommand \url  [0]{\begingroup\@sanitize@url \@url }%
\providecommand \@url [1]{\endgroup\@href {#1}{\urlprefix }}%
\providecommand \urlprefix  [0]{URL }%
\providecommand \Eprint [0]{\href }%
\providecommand \doibase [0]{http://dx.doi.org/}%
\providecommand \selectlanguage [0]{\@gobble}%
\providecommand \bibinfo  [0]{\@secondoftwo}%
\providecommand \bibfield  [0]{\@secondoftwo}%
\providecommand \translation [1]{[#1]}%
\providecommand \BibitemOpen [0]{}%
\providecommand \bibitemStop [0]{}%
\providecommand \bibitemNoStop [0]{.\EOS\space}%
\providecommand \EOS [0]{\spacefactor3000\relax}%
\providecommand \BibitemShut  [1]{\csname bibitem#1\endcsname}%
\let\auto@bib@innerbib\@empty
%</preamble>
\bibitem [{\citenamefont {Tool}(1946)}]{tool1946relation}%
  \BibitemOpen
  \bibfield  {author} {\bibinfo {author} {\bibfnamefont {A.~Q.}\ \bibnamefont
  {Tool}},\ }\bibfield  {title} {\enquote {\bibinfo {title} {Relation between
  inelastic deformability and thermal expansion of glass in its annealing
  range},}\ }\href@noop {} {\bibfield  {journal} {\bibinfo  {journal} {Journal
  of the American Ceramic society}\ }\textbf {\bibinfo {volume} {29}},\
  \bibinfo {pages} {240--253} (\bibinfo {year} {1946})}\BibitemShut {NoStop}%
\bibitem [{\citenamefont {Narayanaswamy}(1971)}]{narayanaswamy1971model}%
  \BibitemOpen
  \bibfield  {author} {\bibinfo {author} {\bibfnamefont {O.}~\bibnamefont
  {Narayanaswamy}},\ }\bibfield  {title} {\enquote {\bibinfo {title} {A model
  of structural relaxation in glass},}\ }\href@noop {} {\bibfield  {journal}
  {\bibinfo  {journal} {Journal of the American Ceramic Society}\ }\textbf
  {\bibinfo {volume} {54}},\ \bibinfo {pages} {491--498} (\bibinfo {year}
  {1971})}\BibitemShut {NoStop}%
\bibitem [{\citenamefont {Micoulaut}(2016)}]{micoulaut2016relaxation}%
  \BibitemOpen
  \bibfield  {author} {\bibinfo {author} {\bibfnamefont {M.}~\bibnamefont
  {Micoulaut}},\ }\bibfield  {title} {\enquote {\bibinfo {title} {Relaxation
  and physical aging in network glasses: a review},}\ }\href@noop {} {\bibfield
   {journal} {\bibinfo  {journal} {Reports on Progress in Physics}\ }\textbf
  {\bibinfo {volume} {79}},\ \bibinfo {pages} {066504} (\bibinfo {year}
  {2016})}\BibitemShut {NoStop}%
\bibitem [{\citenamefont {Leheny}\ and\ \citenamefont
  {Nagel}(1998)}]{leheny_frequency-domain_1998}%
  \BibitemOpen
  \bibfield  {author} {\bibinfo {author} {\bibfnamefont {R.~L.}\ \bibnamefont
  {Leheny}}\ and\ \bibinfo {author} {\bibfnamefont {S.~R.}\ \bibnamefont
  {Nagel}},\ }\bibfield  {title} {\enquote {\bibinfo {title} {Frequency-domain
  study of physical aging in a simple liquid},}\ }\href@noop {} {\bibfield
  {journal} {\bibinfo  {journal} {Physical Review B}\ ,\ \bibinfo {pages} {9}}
  (\bibinfo {year} {1998})}\BibitemShut {NoStop}%
\bibitem [{\citenamefont {Lunkenheimer}\ \emph {et~al.}(2005)\citenamefont
  {Lunkenheimer}, \citenamefont {Wehn}, \citenamefont {Schneider},\ and\
  \citenamefont {Loidl}}]{lunkenheimer_glassy_2005}%
  \BibitemOpen
  \bibfield  {author} {\bibinfo {author} {\bibfnamefont {P.}~\bibnamefont
  {Lunkenheimer}}, \bibinfo {author} {\bibfnamefont {R.}~\bibnamefont {Wehn}},
  \bibinfo {author} {\bibfnamefont {U.}~\bibnamefont {Schneider}}, \ and\
  \bibinfo {author} {\bibfnamefont {A.}~\bibnamefont {Loidl}},\ }\bibfield
  {title} {\enquote {\bibinfo {title} {Glassy {Aging} {Dynamics}},}\ }\href
  {\doibase 10.1103/PhysRevLett.95.055702} {\bibfield  {journal} {\bibinfo
  {journal} {Physical Review Letters}\ }\textbf {\bibinfo {volume} {95}},\
  \bibinfo {pages} {055702} (\bibinfo {year} {2005})}\BibitemShut {NoStop}%
\bibitem [{\citenamefont {Brun}\ \emph {et~al.}(2012)\citenamefont {Brun},
  \citenamefont {Ladieu}, \citenamefont {L’H{\^o}te}, \citenamefont
  {Biroli},\ and\ \citenamefont {Bouchaud}}]{brun2012evidence}%
  \BibitemOpen
  \bibfield  {author} {\bibinfo {author} {\bibfnamefont {C.}~\bibnamefont
  {Brun}}, \bibinfo {author} {\bibfnamefont {F.}~\bibnamefont {Ladieu}},
  \bibinfo {author} {\bibfnamefont {D.}~\bibnamefont {L’H{\^o}te}}, \bibinfo
  {author} {\bibfnamefont {G.}~\bibnamefont {Biroli}}, \ and\ \bibinfo {author}
  {\bibfnamefont {J.}~\bibnamefont {Bouchaud}},\ }\bibfield  {title} {\enquote
  {\bibinfo {title} {Evidence of growing spatial correlations during the aging
  of glassy glycerol},}\ }\href@noop {} {\bibfield  {journal} {\bibinfo
  {journal} {Physical review letters}\ }\textbf {\bibinfo {volume} {109}},\
  \bibinfo {pages} {175702} (\bibinfo {year} {2012})}\BibitemShut {NoStop}%
\bibitem [{\citenamefont {Hecksher}\ \emph {et~al.}(2010)\citenamefont
  {Hecksher}, \citenamefont {Olsen}, \citenamefont {Niss},\ and\ \citenamefont
  {Dyre}}]{hecksher2010physical}%
  \BibitemOpen
  \bibfield  {author} {\bibinfo {author} {\bibfnamefont {T.}~\bibnamefont
  {Hecksher}}, \bibinfo {author} {\bibfnamefont {N.~B.}\ \bibnamefont {Olsen}},
  \bibinfo {author} {\bibfnamefont {K.}~\bibnamefont {Niss}}, \ and\ \bibinfo
  {author} {\bibfnamefont {J.~C.}\ \bibnamefont {Dyre}},\ }\bibfield  {title}
  {\enquote {\bibinfo {title} {Physical aging of molecular glasses studied by a
  device allowing for rapid thermal equilibration},}\ }\href@noop {} {\bibfield
   {journal} {\bibinfo  {journal} {The Journal of chemical physics}\ }\textbf
  {\bibinfo {volume} {133}},\ \bibinfo {pages} {174514} (\bibinfo {year}
  {2010})}\BibitemShut {NoStop}%
\bibitem [{\citenamefont {Hecksher}, \citenamefont {Olsen},\ and\ \citenamefont
  {Dyre}(2015)}]{hecksher2015communication}%
  \BibitemOpen
  \bibfield  {author} {\bibinfo {author} {\bibfnamefont {T.}~\bibnamefont
  {Hecksher}}, \bibinfo {author} {\bibfnamefont {N.~B.}\ \bibnamefont {Olsen}},
  \ and\ \bibinfo {author} {\bibfnamefont {J.~C.}\ \bibnamefont {Dyre}},\
  }\bibfield  {title} {\enquote {\bibinfo {title} {Communication: Direct tests
  of single-parameter aging},}\ }\href@noop {} {\bibfield  {journal} {\bibinfo
  {journal} {The Journal of Chemical Physics}\ }\textbf {\bibinfo {volume}
  {142}},\ \bibinfo {pages} {241103} (\bibinfo {year} {2015})}\BibitemShut
  {NoStop}%
\bibitem [{\citenamefont {Roed}\ \emph {et~al.}(2019)\citenamefont {Roed},
  \citenamefont {Hecksher}, \citenamefont {Dyre},\ and\ \citenamefont
  {Niss}}]{roed2019generalized}%
  \BibitemOpen
  \bibfield  {author} {\bibinfo {author} {\bibfnamefont {L.~A.}\ \bibnamefont
  {Roed}}, \bibinfo {author} {\bibfnamefont {T.}~\bibnamefont {Hecksher}},
  \bibinfo {author} {\bibfnamefont {J.~C.}\ \bibnamefont {Dyre}}, \ and\
  \bibinfo {author} {\bibfnamefont {K.}~\bibnamefont {Niss}},\ }\bibfield
  {title} {\enquote {\bibinfo {title} {Generalized single-parameter aging tests
  and their application to glycerol},}\ }\href@noop {} {\bibfield  {journal}
  {\bibinfo  {journal} {The Journal of Chemical Physics}\ }\textbf {\bibinfo
  {volume} {150}},\ \bibinfo {pages} {044501} (\bibinfo {year}
  {2019})}\BibitemShut {NoStop}%
\bibitem [{\citenamefont {Riechers}\ \emph {et~al.}(2022)\citenamefont
  {Riechers}, \citenamefont {Roed}, \citenamefont {Mehri}, \citenamefont
  {Ingebrigtsen}, \citenamefont {Hecksher}, \citenamefont {Dyre},\ and\
  \citenamefont {Niss}}]{riechers_predicting_2022}%
  \BibitemOpen
  \bibfield  {author} {\bibinfo {author} {\bibfnamefont {B.}~\bibnamefont
  {Riechers}}, \bibinfo {author} {\bibfnamefont {L.~A.}\ \bibnamefont {Roed}},
  \bibinfo {author} {\bibfnamefont {S.}~\bibnamefont {Mehri}}, \bibinfo
  {author} {\bibfnamefont {T.~S.}\ \bibnamefont {Ingebrigtsen}}, \bibinfo
  {author} {\bibfnamefont {T.}~\bibnamefont {Hecksher}}, \bibinfo {author}
  {\bibfnamefont {J.~C.}\ \bibnamefont {Dyre}}, \ and\ \bibinfo {author}
  {\bibfnamefont {K.}~\bibnamefont {Niss}},\ }\bibfield  {title} {\enquote
  {\bibinfo {title} {Predicting nonlinear physical aging of glasses from
  equilibrium relaxation via the material time},}\ }\href@noop {} {\bibfield
  {journal} {\bibinfo  {journal} {Science advances}\ }\textbf {\bibinfo
  {volume} {8}},\ \bibinfo {pages} {eabl9809} (\bibinfo {year}
  {2022})}\BibitemShut {NoStop}%
\bibitem [{\citenamefont {Kovacs}(1958)}]{kovacs1958contraction}%
  \BibitemOpen
  \bibfield  {author} {\bibinfo {author} {\bibfnamefont {A.~J.}\ \bibnamefont
  {Kovacs}},\ }\bibfield  {title} {\enquote {\bibinfo {title} {La contraction
  isotherme du volume des polym{\`e}res amorphes},}\ }\href@noop {} {\bibfield
  {journal} {\bibinfo  {journal} {Journal of polymer science}\ }\textbf
  {\bibinfo {volume} {30}},\ \bibinfo {pages} {131--147} (\bibinfo {year}
  {1958})}\BibitemShut {NoStop}%
\bibitem [{\citenamefont {McKenna}(1995)}]{mckenna1995physics}%
  \BibitemOpen
  \bibfield  {author} {\bibinfo {author} {\bibfnamefont {G.~B.}\ \bibnamefont
  {McKenna}},\ }\bibfield  {title} {\enquote {\bibinfo {title} {On the physics
  required for prediction of long term performance of polymers and their
  composites},}\ }\href@noop {} {\bibfield  {journal} {\bibinfo  {journal}
  {Keynote lectures in selected topics of polymer science}\ ,\ \bibinfo {pages}
  {139}} (\bibinfo {year} {1995})}\BibitemShut {NoStop}%
\bibitem [{\citenamefont {Abou}, \citenamefont {Bonn},\ and\ \citenamefont
  {Meunier}(2001)}]{abou2001aging}%
  \BibitemOpen
  \bibfield  {author} {\bibinfo {author} {\bibfnamefont {B.}~\bibnamefont
  {Abou}}, \bibinfo {author} {\bibfnamefont {D.}~\bibnamefont {Bonn}}, \ and\
  \bibinfo {author} {\bibfnamefont {J.}~\bibnamefont {Meunier}},\ }\bibfield
  {title} {\enquote {\bibinfo {title} {Aging dynamics in a colloidal glass},}\
  }\href@noop {} {\bibfield  {journal} {\bibinfo  {journal} {Physical Review
  E}\ }\textbf {\bibinfo {volume} {64}},\ \bibinfo {pages} {021510} (\bibinfo
  {year} {2001})}\BibitemShut {NoStop}%
\bibitem [{\citenamefont {Richard}\ \emph {et~al.}(2005)\citenamefont
  {Richard}, \citenamefont {Nicodemi}, \citenamefont {Delannay}, \citenamefont
  {Ribiere},\ and\ \citenamefont {Bideau}}]{richard2005slow}%
  \BibitemOpen
  \bibfield  {author} {\bibinfo {author} {\bibfnamefont {P.}~\bibnamefont
  {Richard}}, \bibinfo {author} {\bibfnamefont {M.}~\bibnamefont {Nicodemi}},
  \bibinfo {author} {\bibfnamefont {R.}~\bibnamefont {Delannay}}, \bibinfo
  {author} {\bibfnamefont {P.}~\bibnamefont {Ribiere}}, \ and\ \bibinfo
  {author} {\bibfnamefont {D.}~\bibnamefont {Bideau}},\ }\bibfield  {title}
  {\enquote {\bibinfo {title} {Slow relaxation and compaction of granular
  systems},}\ }\href@noop {} {\bibfield  {journal} {\bibinfo  {journal} {Nature
  materials}\ }\textbf {\bibinfo {volume} {4}},\ \bibinfo {pages} {121--128}
  (\bibinfo {year} {2005})}\BibitemShut {NoStop}%
\bibitem [{\citenamefont {Mandal}\ and\ \citenamefont
  {Sollich}(2020)}]{mandal2020multiple}%
  \BibitemOpen
  \bibfield  {author} {\bibinfo {author} {\bibfnamefont {R.}~\bibnamefont
  {Mandal}}\ and\ \bibinfo {author} {\bibfnamefont {P.}~\bibnamefont
  {Sollich}},\ }\bibfield  {title} {\enquote {\bibinfo {title} {Multiple types
  of aging in active glasses},}\ }\href@noop {} {\bibfield  {journal} {\bibinfo
   {journal} {Physical Review Letters}\ }\textbf {\bibinfo {volume} {125}},\
  \bibinfo {pages} {218001} (\bibinfo {year} {2020})}\BibitemShut {NoStop}%
\bibitem [{\citenamefont {Andersen}\ \emph {et~al.}(2019)\citenamefont
  {Andersen}, \citenamefont {Mikkelsen}, \citenamefont {Kristiansen},\ and\
  \citenamefont {Hinge}}]{andersen2019accelerated}%
  \BibitemOpen
  \bibfield  {author} {\bibinfo {author} {\bibfnamefont {E.}~\bibnamefont
  {Andersen}}, \bibinfo {author} {\bibfnamefont {R.}~\bibnamefont {Mikkelsen}},
  \bibinfo {author} {\bibfnamefont {S.}~\bibnamefont {Kristiansen}}, \ and\
  \bibinfo {author} {\bibfnamefont {M.}~\bibnamefont {Hinge}},\ }\bibfield
  {title} {\enquote {\bibinfo {title} {Accelerated physical ageing of poly (1,
  4-cyclohexylenedimethylene-co-2, 2, 4, 4-tetramethyl-1, 3-cyclobutanediol
  terephthalate)},}\ }\href@noop {} {\bibfield  {journal} {\bibinfo  {journal}
  {RSC advances}\ }\textbf {\bibinfo {volume} {9}},\ \bibinfo {pages}
  {14209--14219} (\bibinfo {year} {2019})}\BibitemShut {NoStop}%
\bibitem [{\citenamefont {Jena}\ \emph {et~al.}(2019)\citenamefont {Jena},
  \citenamefont {Tokas}, \citenamefont {Thakur},\ and\ \citenamefont
  {Udupa}}]{jena2019study}%
  \BibitemOpen
  \bibfield  {author} {\bibinfo {author} {\bibfnamefont {S.}~\bibnamefont
  {Jena}}, \bibinfo {author} {\bibfnamefont {R.}~\bibnamefont {Tokas}},
  \bibinfo {author} {\bibfnamefont {S.}~\bibnamefont {Thakur}}, \ and\ \bibinfo
  {author} {\bibfnamefont {D.}~\bibnamefont {Udupa}},\ }\bibfield  {title}
  {\enquote {\bibinfo {title} {Study of aging effects on optical properties and
  residual stress of hfo2 thin film},}\ }\href@noop {} {\bibfield  {journal}
  {\bibinfo  {journal} {Optik}\ }\textbf {\bibinfo {volume} {185}},\ \bibinfo
  {pages} {71--81} (\bibinfo {year} {2019})}\BibitemShut {NoStop}%
\bibitem [{\citenamefont {Arceri}\ \emph {et~al.}(2020)\citenamefont {Arceri},
  \citenamefont {Landes}, \citenamefont {Berthier},\ and\ \citenamefont
  {Biroli}}]{arceri2020glasses}%
  \BibitemOpen
  \bibfield  {author} {\bibinfo {author} {\bibfnamefont {F.}~\bibnamefont
  {Arceri}}, \bibinfo {author} {\bibfnamefont {F.~P.}\ \bibnamefont {Landes}},
  \bibinfo {author} {\bibfnamefont {L.}~\bibnamefont {Berthier}}, \ and\
  \bibinfo {author} {\bibfnamefont {G.}~\bibnamefont {Biroli}},\ }\bibfield
  {title} {\enquote {\bibinfo {title} {Glasses and aging: A statistical
  mechanics perspective},}\ }\href@noop {} {\bibfield  {journal} {\bibinfo
  {journal} {arXiv preprint arXiv:2006.09725}\ } (\bibinfo {year}
  {2020})}\BibitemShut {NoStop}%
\bibitem [{\citenamefont {Lillie}(1933)}]{lillie1933viscosity}%
  \BibitemOpen
  \bibfield  {author} {\bibinfo {author} {\bibfnamefont {H.~R.}\ \bibnamefont
  {Lillie}},\ }\bibfield  {title} {\enquote {\bibinfo {title}
  {Viscosity-time-temperature relations in glass at annealing temperatures},}\
  }\href@noop {} {\bibfield  {journal} {\bibinfo  {journal} {Journal of the
  American Ceramic Society}\ }\textbf {\bibinfo {volume} {16}},\ \bibinfo
  {pages} {619--631} (\bibinfo {year} {1933})}\BibitemShut {NoStop}%
\bibitem [{\citenamefont {Kovacs}(1964)}]{kovacs1964transition}%
  \BibitemOpen
  \bibfield  {author} {\bibinfo {author} {\bibfnamefont {A.~J.}\ \bibnamefont
  {Kovacs}},\ }\bibfield  {title} {\enquote {\bibinfo {title} {Transition
  vitreuse dans les polym{\`e}res amorphes. etude ph{\'e}nom{\'e}nologique},}\
  }in\ \href@noop {} {\emph {\bibinfo {booktitle} {Fortschritte der
  hochpolymeren-forschung}}}\ (\bibinfo  {publisher} {Springer},\ \bibinfo
  {year} {1964})\ pp.\ \bibinfo {pages} {394--507}\BibitemShut {NoStop}%
\bibitem [{\citenamefont {Moynihan}\ \emph {et~al.}(1976)\citenamefont
  {Moynihan}, \citenamefont {Macedo}, \citenamefont {Montrose}, \citenamefont
  {Montrose}, \citenamefont {Gupta}, \citenamefont {DeBolt}, \citenamefont
  {Dill}, \citenamefont {Dom}, \citenamefont {Drake}, \citenamefont {Easteal}
  \emph {et~al.}}]{moynihan1976structural}%
  \BibitemOpen
  \bibfield  {author} {\bibinfo {author} {\bibfnamefont {C.}~\bibnamefont
  {Moynihan}}, \bibinfo {author} {\bibfnamefont {P.}~\bibnamefont {Macedo}},
  \bibinfo {author} {\bibfnamefont {C.}~\bibnamefont {Montrose}}, \bibinfo
  {author} {\bibfnamefont {C.}~\bibnamefont {Montrose}}, \bibinfo {author}
  {\bibfnamefont {P.}~\bibnamefont {Gupta}}, \bibinfo {author} {\bibfnamefont
  {M.}~\bibnamefont {DeBolt}}, \bibinfo {author} {\bibfnamefont
  {J.}~\bibnamefont {Dill}}, \bibinfo {author} {\bibfnamefont {B.}~\bibnamefont
  {Dom}}, \bibinfo {author} {\bibfnamefont {P.}~\bibnamefont {Drake}}, \bibinfo
  {author} {\bibfnamefont {A.}~\bibnamefont {Easteal}},  \emph {et~al.},\
  }\bibfield  {title} {\enquote {\bibinfo {title} {Structural relaxation in
  vitreous materials},}\ }\href@noop {} {\bibfield  {journal} {\bibinfo
  {journal} {Annals of the New York Academy of Sciences}\ }\textbf {\bibinfo
  {volume} {279}},\ \bibinfo {pages} {15--35} (\bibinfo {year}
  {1976})}\BibitemShut {NoStop}%
\bibitem [{\citenamefont {Mauro}, \citenamefont {Loucks},\ and\ \citenamefont
  {Gupta}(2009)}]{mauro2009fictive}%
  \BibitemOpen
  \bibfield  {author} {\bibinfo {author} {\bibfnamefont {J.~C.}\ \bibnamefont
  {Mauro}}, \bibinfo {author} {\bibfnamefont {R.~J.}\ \bibnamefont {Loucks}}, \
  and\ \bibinfo {author} {\bibfnamefont {P.~K.}\ \bibnamefont {Gupta}},\
  }\bibfield  {title} {\enquote {\bibinfo {title} {Fictive temperature and the
  glassy state},}\ }\href@noop {} {\bibfield  {journal} {\bibinfo  {journal}
  {Journal of the American Ceramic Society}\ }\textbf {\bibinfo {volume}
  {92}},\ \bibinfo {pages} {75--86} (\bibinfo {year} {2009})}\BibitemShut
  {NoStop}%
\bibitem [{\citenamefont {Kovacs}\ \emph {et~al.}(1979)\citenamefont {Kovacs},
  \citenamefont {Aklonis}, \citenamefont {Hutchinson},\ and\ \citenamefont
  {Ramos}}]{kovacs1979isobaric}%
  \BibitemOpen
  \bibfield  {author} {\bibinfo {author} {\bibfnamefont {A.~J.}\ \bibnamefont
  {Kovacs}}, \bibinfo {author} {\bibfnamefont {J.~J.}\ \bibnamefont {Aklonis}},
  \bibinfo {author} {\bibfnamefont {J.~M.}\ \bibnamefont {Hutchinson}}, \ and\
  \bibinfo {author} {\bibfnamefont {A.~R.}\ \bibnamefont {Ramos}},\ }\bibfield
  {title} {\enquote {\bibinfo {title} {Isobaric volume and enthalpy recovery of
  glasses. ii. a transparent multiparameter theory},}\ }\href@noop {}
  {\bibfield  {journal} {\bibinfo  {journal} {Journal of Polymer Science:
  Polymer Physics Edition}\ }\textbf {\bibinfo {volume} {17}},\ \bibinfo
  {pages} {1097--1162} (\bibinfo {year} {1979})}\BibitemShut {NoStop}%
\bibitem [{\citenamefont {Mazinani}\ and\ \citenamefont
  {Richert}(2012)}]{mazinani2012enthalpy}%
  \BibitemOpen
  \bibfield  {author} {\bibinfo {author} {\bibfnamefont {S.~K.}\ \bibnamefont
  {Mazinani}}\ and\ \bibinfo {author} {\bibfnamefont {R.}~\bibnamefont
  {Richert}},\ }\bibfield  {title} {\enquote {\bibinfo {title} {Enthalpy
  recovery in glassy materials: Heterogeneous versus homogenous models},}\
  }\href@noop {} {\bibfield  {journal} {\bibinfo  {journal} {The Journal of
  Chemical Physics}\ }\textbf {\bibinfo {volume} {136}},\ \bibinfo {pages}
  {174515} (\bibinfo {year} {2012})}\BibitemShut {NoStop}%
\bibitem [{\citenamefont {Richert}\ \emph {et~al.}(2013)\citenamefont
  {Richert}, \citenamefont {Lunkenheimer}, \citenamefont {Kastner},\ and\
  \citenamefont {Loidl}}]{richert_derivation_2013}%
  \BibitemOpen
  \bibfield  {author} {\bibinfo {author} {\bibfnamefont {R.}~\bibnamefont
  {Richert}}, \bibinfo {author} {\bibfnamefont {P.}~\bibnamefont
  {Lunkenheimer}}, \bibinfo {author} {\bibfnamefont {S.}~\bibnamefont
  {Kastner}}, \ and\ \bibinfo {author} {\bibfnamefont {A.}~\bibnamefont
  {Loidl}},\ }\bibfield  {title} {\enquote {\bibinfo {title} {On the
  {Derivation} of {Equilibrium} {Relaxation} {Times} from {Aging}
  {Experiments}},}\ }\href {\doibase 10.1021/jp311149n} {\bibfield  {journal}
  {\bibinfo  {journal} {The Journal of Physical Chemistry B}\ }\textbf
  {\bibinfo {volume} {117}},\ \bibinfo {pages} {12689--12694} (\bibinfo {year}
  {2013})}\BibitemShut {NoStop}%
\bibitem [{\citenamefont {Sep{\'u}lveda}\ \emph {et~al.}(2014)\citenamefont
  {Sep{\'u}lveda}, \citenamefont {Tylinski}, \citenamefont {Guiseppi-Elie},
  \citenamefont {Richert},\ and\ \citenamefont {Ediger}}]{sepulveda2014role}%
  \BibitemOpen
  \bibfield  {author} {\bibinfo {author} {\bibfnamefont {A.}~\bibnamefont
  {Sep{\'u}lveda}}, \bibinfo {author} {\bibfnamefont {M.}~\bibnamefont
  {Tylinski}}, \bibinfo {author} {\bibfnamefont {A.}~\bibnamefont
  {Guiseppi-Elie}}, \bibinfo {author} {\bibfnamefont {R.}~\bibnamefont
  {Richert}}, \ and\ \bibinfo {author} {\bibfnamefont {M.}~\bibnamefont
  {Ediger}},\ }\bibfield  {title} {\enquote {\bibinfo {title} {Role of
  fragility in the formation of highly stable organic glasses},}\ }\href@noop
  {} {\bibfield  {journal} {\bibinfo  {journal} {Physical review letters}\
  }\textbf {\bibinfo {volume} {113}},\ \bibinfo {pages} {045901} (\bibinfo
  {year} {2014})}\BibitemShut {NoStop}%
\bibitem [{\citenamefont {Vila-Costa}\ \emph {et~al.}(2020)\citenamefont
  {Vila-Costa}, \citenamefont {R{\`a}fols-Rib{\'e}}, \citenamefont
  {Gonz{\'a}lez-Silveira}, \citenamefont {Lopeandia}, \citenamefont
  {Abad-Mu{\~n}oz},\ and\ \citenamefont
  {Rodr{\'\i}guez-Viejo}}]{vila2020nucleation}%
  \BibitemOpen
  \bibfield  {author} {\bibinfo {author} {\bibfnamefont {A.}~\bibnamefont
  {Vila-Costa}}, \bibinfo {author} {\bibfnamefont {J.}~\bibnamefont
  {R{\`a}fols-Rib{\'e}}}, \bibinfo {author} {\bibfnamefont {M.}~\bibnamefont
  {Gonz{\'a}lez-Silveira}}, \bibinfo {author} {\bibfnamefont {A.}~\bibnamefont
  {Lopeandia}}, \bibinfo {author} {\bibfnamefont {L.}~\bibnamefont
  {Abad-Mu{\~n}oz}}, \ and\ \bibinfo {author} {\bibfnamefont {J.}~\bibnamefont
  {Rodr{\'\i}guez-Viejo}},\ }\bibfield  {title} {\enquote {\bibinfo {title}
  {Nucleation and growth of the supercooled liquid phase control glass
  transition in bulk ultrastable glasses},}\ }\href@noop {} {\bibfield
  {journal} {\bibinfo  {journal} {Physical Review Letters}\ }\textbf {\bibinfo
  {volume} {124}},\ \bibinfo {pages} {076002} (\bibinfo {year}
  {2020})}\BibitemShut {NoStop}%
\bibitem [{\citenamefont {Vila-Costa}\ \emph {et~al.}(2023)\citenamefont
  {Vila-Costa}, \citenamefont {Gonzalez-Silveira}, \citenamefont
  {Rodr{\'\i}guez-Tinoco}, \citenamefont {Rodr{\'\i}guez-L{\'o}pez},\ and\
  \citenamefont {Rodriguez-Viejo}}]{vila2023emergence}%
  \BibitemOpen
  \bibfield  {author} {\bibinfo {author} {\bibfnamefont {A.}~\bibnamefont
  {Vila-Costa}}, \bibinfo {author} {\bibfnamefont {M.}~\bibnamefont
  {Gonzalez-Silveira}}, \bibinfo {author} {\bibfnamefont {C.}~\bibnamefont
  {Rodr{\'\i}guez-Tinoco}}, \bibinfo {author} {\bibfnamefont {M.}~\bibnamefont
  {Rodr{\'\i}guez-L{\'o}pez}}, \ and\ \bibinfo {author} {\bibfnamefont
  {J.}~\bibnamefont {Rodriguez-Viejo}},\ }\bibfield  {title} {\enquote
  {\bibinfo {title} {Emergence of equilibrated liquid regions within the
  glass},}\ }\href@noop {} {\bibfield  {journal} {\bibinfo  {journal} {Nature
  Physics}\ }\textbf {\bibinfo {volume} {19}},\ \bibinfo {pages} {114--119}
  (\bibinfo {year} {2023})}\BibitemShut {NoStop}%
\bibitem [{\citenamefont {Hecht}(2012)}]{freefem}%
  \BibitemOpen
  \bibfield  {author} {\bibinfo {author} {\bibfnamefont {F.}~\bibnamefont
  {Hecht}},\ }\bibfield  {title} {\enquote {\bibinfo {title} {New development
  in freefem++},}\ }\href@noop {} {\bibfield  {journal} {\bibinfo  {journal}
  {J. Numer. Math.}\ }\textbf {\bibinfo {volume} {20}},\ \bibinfo {pages}
  {251--265} (\bibinfo {year} {2012})}\BibitemShut {NoStop}%
\bibitem [{\citenamefont {Astrath}\ \emph {et~al.}(2005)\citenamefont
  {Astrath}, \citenamefont {Rohling}, \citenamefont {Medina}, \citenamefont
  {Bento}, \citenamefont {Baesso}, \citenamefont {Jacinto}, \citenamefont
  {Catunda}, \citenamefont {Lima}, \citenamefont {Gandra}, \citenamefont
  {Bell},\ and\ \citenamefont {Anjos}}]{Astrath_2005}%
  \BibitemOpen
  \bibfield  {author} {\bibinfo {author} {\bibfnamefont {N.}~\bibnamefont
  {Astrath}}, \bibinfo {author} {\bibfnamefont {J.}~\bibnamefont {Rohling}},
  \bibinfo {author} {\bibfnamefont {A.}~\bibnamefont {Medina}}, \bibinfo
  {author} {\bibfnamefont {A.}~\bibnamefont {Bento}}, \bibinfo {author}
  {\bibfnamefont {M.}~\bibnamefont {Baesso}}, \bibinfo {author} {\bibfnamefont
  {C.}~\bibnamefont {Jacinto}}, \bibinfo {author} {\bibfnamefont
  {T.}~\bibnamefont {Catunda}}, \bibinfo {author} {\bibfnamefont
  {S.}~\bibnamefont {Lima}}, \bibinfo {author} {\bibfnamefont {F.}~\bibnamefont
  {Gandra}}, \bibinfo {author} {\bibfnamefont {M.~J.}\ \bibnamefont {Bell}}, \
  and\ \bibinfo {author} {\bibfnamefont {V.}~\bibnamefont {Anjos}},\ }\bibfield
   {title} {\enquote {\bibinfo {title} {Time-resolved thermal lens measurements
  of the thermo-optical properties of glasses at low temperature down to 20
  k},}\ }\href@noop {} {\bibfield  {journal} {\bibinfo  {journal} {Physical
  Review B - Condensed Matter and Materials Physics}\ }\textbf {\bibinfo
  {volume} {71}} (\bibinfo {year} {2005})}\BibitemShut {NoStop}%
\bibitem [{\citenamefont {Birge}(1986)}]{PhysRevB.34.1631}%
  \BibitemOpen
  \bibfield  {author} {\bibinfo {author} {\bibfnamefont {N.~O.}\ \bibnamefont
  {Birge}},\ }\bibfield  {title} {\enquote {\bibinfo {title} {Specific-heat
  spectroscopy of glycerol and propylene glycol near the glass transition},}\
  }\href {\doibase 10.1103/PhysRevB.34.1631} {\bibfield  {journal} {\bibinfo
  {journal} {Phys. Rev. B}\ }\textbf {\bibinfo {volume} {34}},\ \bibinfo
  {pages} {1631--1642} (\bibinfo {year} {1986})}\BibitemShut {NoStop}%
\bibitem [{\citenamefont {Minakov}, \citenamefont {Adamovsky},\ and\
  \citenamefont {Schick}(2003)}]{minakov2003advanced}%
  \BibitemOpen
  \bibfield  {author} {\bibinfo {author} {\bibfnamefont {A.}~\bibnamefont
  {Minakov}}, \bibinfo {author} {\bibfnamefont {S.}~\bibnamefont {Adamovsky}},
  \ and\ \bibinfo {author} {\bibfnamefont {C.}~\bibnamefont {Schick}},\
  }\bibfield  {title} {\enquote {\bibinfo {title} {Advanced two-channel ac
  calorimeter for simultaneous measurements of complex heat capacity and
  complex thermal conductivity},}\ }\href@noop {} {\bibfield  {journal}
  {\bibinfo  {journal} {Thermochimica acta}\ }\textbf {\bibinfo {volume}
  {403}},\ \bibinfo {pages} {89--103} (\bibinfo {year} {2003})}\BibitemShut
  {NoStop}%
\bibitem [{\citenamefont {Mazurin}(1977)}]{mazurin1977relaxation}%
  \BibitemOpen
  \bibfield  {author} {\bibinfo {author} {\bibfnamefont {O.}~\bibnamefont
  {Mazurin}},\ }\bibfield  {title} {\enquote {\bibinfo {title} {Relaxation
  phenomena in glass},}\ }\href@noop {} {\bibfield  {journal} {\bibinfo
  {journal} {Journal of Non-Crystalline Solids}\ }\textbf {\bibinfo {volume}
  {25}},\ \bibinfo {pages} {129--169} (\bibinfo {year} {1977})}\BibitemShut
  {NoStop}%
\bibitem [{\citenamefont {Hodge}(1994)}]{hodge1994enthalpy}%
  \BibitemOpen
  \bibfield  {author} {\bibinfo {author} {\bibfnamefont {I.~M.}\ \bibnamefont
  {Hodge}},\ }\bibfield  {title} {\enquote {\bibinfo {title} {Enthalpy
  relaxation and recovery in amorphous materials},}\ }\href@noop {} {\bibfield
  {journal} {\bibinfo  {journal} {Journal of Non-Crystalline Solids}\ }\textbf
  {\bibinfo {volume} {169}},\ \bibinfo {pages} {211--266} (\bibinfo {year}
  {1994})}\BibitemShut {NoStop}%
\bibitem [{\citenamefont {Mauro}, \citenamefont {Allan},\ and\ \citenamefont
  {Potuzak}(2009)}]{mauro2009nonequilibrium}%
  \BibitemOpen
  \bibfield  {author} {\bibinfo {author} {\bibfnamefont {J.~C.}\ \bibnamefont
  {Mauro}}, \bibinfo {author} {\bibfnamefont {D.~C.}\ \bibnamefont {Allan}}, \
  and\ \bibinfo {author} {\bibfnamefont {M.}~\bibnamefont {Potuzak}},\
  }\bibfield  {title} {\enquote {\bibinfo {title} {Nonequilibrium viscosity of
  glass},}\ }\href@noop {} {\bibfield  {journal} {\bibinfo  {journal} {Physical
  Review B}\ }\textbf {\bibinfo {volume} {80}},\ \bibinfo {pages} {094204}
  (\bibinfo {year} {2009})}\BibitemShut {NoStop}%
\bibitem [{\citenamefont {Lunkenheimer}\ and\ \citenamefont
  {Loidl}(2002)}]{lunkenheimer_dielectric_2002}%
  \BibitemOpen
  \bibfield  {author} {\bibinfo {author} {\bibfnamefont {P.}~\bibnamefont
  {Lunkenheimer}}\ and\ \bibinfo {author} {\bibfnamefont {A.}~\bibnamefont
  {Loidl}},\ }\bibfield  {title} {\enquote {\bibinfo {title} {Dielectric
  spectroscopy of glass-forming materials: a-relaxation and excess wing},}\
  }\href@noop {} {\bibfield  {journal} {\bibinfo  {journal} {Chemical Physics}\
  ,\ \bibinfo {pages} {15}} (\bibinfo {year} {2002})}\BibitemShut {NoStop}%
\bibitem [{\citenamefont {Mazurin}, \citenamefont {Kluyev},\ and\ \citenamefont
  {Stolyar}(1983)}]{mazurin1983temperature}%
  \BibitemOpen
  \bibfield  {author} {\bibinfo {author} {\bibfnamefont {O.}~\bibnamefont
  {Mazurin}}, \bibinfo {author} {\bibfnamefont {V.}~\bibnamefont {Kluyev}}, \
  and\ \bibinfo {author} {\bibfnamefont {S.}~\bibnamefont {Stolyar}},\
  }\bibfield  {title} {\enquote {\bibinfo {title} {Temperature dependences of
  structural relaxation times at constant fictive temperatures in oxide
  glasses},}\ }\href@noop {} {\bibfield  {journal} {\bibinfo  {journal}
  {Glastech Ber}\ }\textbf {\bibinfo {volume} {56}},\ \bibinfo {pages}
  {1148--1153} (\bibinfo {year} {1983})}\BibitemShut {NoStop}%
\bibitem [{\citenamefont {Alvarez}\ and\ \citenamefont
  {Alegria}(1991)}]{alvarez1991colmenero}%
  \BibitemOpen
  \bibfield  {author} {\bibinfo {author} {\bibfnamefont {F.}~\bibnamefont
  {Alvarez}}\ and\ \bibinfo {author} {\bibfnamefont {A.}~\bibnamefont
  {Alegria}},\ }\bibfield  {title} {\enquote {\bibinfo {title} {Colmenero,
  relationship between the time-domain kohlrausch-williams-watts and
  frequency-domain havriliak-negami relaxation functions},}\ }\href@noop {}
  {\bibfield  {journal} {\bibinfo  {journal} {J. Phys. Rev. B}\ }\textbf
  {\bibinfo {volume} {44}},\ \bibinfo {pages} {7306} (\bibinfo {year}
  {1991})}\BibitemShut {NoStop}%
\bibitem [{\citenamefont {Moynihan}, \citenamefont {Crichton},\ and\
  \citenamefont {Opalka}(1991)}]{moynihan1991linear}%
  \BibitemOpen
  \bibfield  {author} {\bibinfo {author} {\bibfnamefont {C.}~\bibnamefont
  {Moynihan}}, \bibinfo {author} {\bibfnamefont {S.}~\bibnamefont {Crichton}},
  \ and\ \bibinfo {author} {\bibfnamefont {S.}~\bibnamefont {Opalka}},\
  }\bibfield  {title} {\enquote {\bibinfo {title} {Linear and non-linear
  structural relaxation},}\ }\href@noop {} {\bibfield  {journal} {\bibinfo
  {journal} {Journal of non-crystalline solids}\ }\textbf {\bibinfo {volume}
  {131}},\ \bibinfo {pages} {420--434} (\bibinfo {year} {1991})}\BibitemShut
  {NoStop}%
\bibitem [{\citenamefont {Lubchenko}\ and\ \citenamefont
  {Wolynes}(2004)}]{lubchenko2004theory}%
  \BibitemOpen
  \bibfield  {author} {\bibinfo {author} {\bibfnamefont {V.}~\bibnamefont
  {Lubchenko}}\ and\ \bibinfo {author} {\bibfnamefont {P.~G.}\ \bibnamefont
  {Wolynes}},\ }\bibfield  {title} {\enquote {\bibinfo {title} {Theory of aging
  in structural glasses},}\ }\href@noop {} {\bibfield  {journal} {\bibinfo
  {journal} {The Journal of chemical physics}\ }\textbf {\bibinfo {volume}
  {121}},\ \bibinfo {pages} {2852--2865} (\bibinfo {year} {2004})}\BibitemShut
  {NoStop}%
\bibitem [{\citenamefont {Ozawa}\ \emph {et~al.}(2018)\citenamefont {Ozawa},
  \citenamefont {Berthier}, \citenamefont {Biroli}, \citenamefont {Rosso},\
  and\ \citenamefont {Tarjus}}]{ozawa2018random}%
  \BibitemOpen
  \bibfield  {author} {\bibinfo {author} {\bibfnamefont {M.}~\bibnamefont
  {Ozawa}}, \bibinfo {author} {\bibfnamefont {L.}~\bibnamefont {Berthier}},
  \bibinfo {author} {\bibfnamefont {G.}~\bibnamefont {Biroli}}, \bibinfo
  {author} {\bibfnamefont {A.}~\bibnamefont {Rosso}}, \ and\ \bibinfo {author}
  {\bibfnamefont {G.}~\bibnamefont {Tarjus}},\ }\bibfield  {title} {\enquote
  {\bibinfo {title} {Random critical point separates brittle and ductile
  yielding transitions in amorphous materials},}\ }\href@noop {} {\bibfield
  {journal} {\bibinfo  {journal} {Proceedings of the National Academy of
  Sciences}\ }\textbf {\bibinfo {volume} {115}},\ \bibinfo {pages} {6656--6661}
  (\bibinfo {year} {2018})}\BibitemShut {NoStop}%
\bibitem [{Note()}]{Note}%
  \BibitemOpen
  Note,\ \href@noop {} {}\bibinfo {note} {One of the arguments put forward by
  ref~\cite {albert2021searching} to explain the absence of a Gardner
  transition in glycerol is the generally well-annealed character of structural
  glasses (compared for example to colloids or granular systems). It would thus
  be interesting to search for Gardner transition in a poorly annealed (\textit
  {i. e.} unstable) glass.}\BibitemShut {Stop}%
\bibitem [{\citenamefont {Albert}\ \emph {et~al.}(2021)\citenamefont {Albert},
  \citenamefont {Biroli}, \citenamefont {Ladieu}, \citenamefont {Tourbot},\
  and\ \citenamefont {Urbani}}]{albert2021searching}%
  \BibitemOpen
  \bibfield  {author} {\bibinfo {author} {\bibfnamefont {S.}~\bibnamefont
  {Albert}}, \bibinfo {author} {\bibfnamefont {G.}~\bibnamefont {Biroli}},
  \bibinfo {author} {\bibfnamefont {F.}~\bibnamefont {Ladieu}}, \bibinfo
  {author} {\bibfnamefont {R.}~\bibnamefont {Tourbot}}, \ and\ \bibinfo
  {author} {\bibfnamefont {P.}~\bibnamefont {Urbani}},\ }\bibfield  {title}
  {\enquote {\bibinfo {title} {Searching for the gardner transition in glassy
  glycerol},}\ }\href@noop {} {\bibfield  {journal} {\bibinfo  {journal}
  {Physical Review Letters}\ }\textbf {\bibinfo {volume} {126}},\ \bibinfo
  {pages} {028001} (\bibinfo {year} {2021})}\BibitemShut {NoStop}%
  \bibitem [{Pol()}]{Polymer_database_Dielectric}%
  \BibitemOpen
  \href@noop {} {\enquote {\bibinfo {title} {Polymer database, dielectric
  constants},}\ }\bibinfo {howpublished}
  {\url{https://polymerdatabase.com/polymer physics/Epsilon Table.html}},\
  \bibinfo {note} {accessed: 2022-10-04}\BibitemShut {NoStop}%
\bibitem [{\citenamefont {Davidson}\ and\ \citenamefont
  {Cole}(1951)}]{davidson1951dielectric}%
  \BibitemOpen
  \bibfield  {author} {\bibinfo {author} {\bibfnamefont {D.~W.}\ \bibnamefont
  {Davidson}}\ and\ \bibinfo {author} {\bibfnamefont {R.~H.}\ \bibnamefont
  {Cole}},\ }\bibfield  {title} {\enquote {\bibinfo {title} {Dielectric
  relaxation in glycerol, propylene glycol, and n-propanol},}\ }\href@noop {}
  {\bibfield  {journal} {\bibinfo  {journal} {The Journal of Chemical Physics}\
  }\textbf {\bibinfo {volume} {19}},\ \bibinfo {pages} {1484--1490} (\bibinfo
  {year} {1951})}\BibitemShut {NoStop}%
\bibitem [{\citenamefont {Ryabov}\ \emph {et~al.}(2003)\citenamefont {Ryabov},
  \citenamefont {Hayashi}, \citenamefont {Gutina},\ and\ \citenamefont
  {Feldman}}]{ryabov2003features}%
  \BibitemOpen
  \bibfield  {author} {\bibinfo {author} {\bibfnamefont {Y.~E.}\ \bibnamefont
  {Ryabov}}, \bibinfo {author} {\bibfnamefont {Y.}~\bibnamefont {Hayashi}},
  \bibinfo {author} {\bibfnamefont {A.}~\bibnamefont {Gutina}}, \ and\ \bibinfo
  {author} {\bibfnamefont {Y.}~\bibnamefont {Feldman}},\ }\bibfield  {title}
  {\enquote {\bibinfo {title} {Features of supercooled glycerol dynamics},}\
  }\href@noop {} {\bibfield  {journal} {\bibinfo  {journal} {Physical Review
  B}\ }\textbf {\bibinfo {volume} {67}},\ \bibinfo {pages} {132202} (\bibinfo
  {year} {2003})}\BibitemShut {NoStop}%
\end{thebibliography}

%%%%%%%%%%%%%%%%%%%%%%%%%%%%%%%%%%%%%%%%%%%%%%%%%%%%%%%%%%%%%%%%%%%%%%%%%%%%%%%%%%%%%%%
%%%%%%%%%%%%%%%%%%%%%%%%%%%%%%%%%%%%%%%%%%%%%%%%%%%%%%%%%%%%%%%%%%%%%%%%%%%%%%%%%%%%%%%
\clearpage
\onecolumngrid
\renewcommand\thefigure{S\arabic{figure}}    
\renewcommand\thetable{S\arabic{table}}    

\renewcommand{\theequation}{S\arabic{equation}}
\setcounter{equation}{0}
\setcounter{figure}{0}
\setcounter{table}{0}
\setcounter{section}{0}

%\maintextlabel{fig1}{1}
%\maintextlabel{fig2}{2}
%\maintextlabel{fig3}{3}
%\maintextlabel{fig4}{4}
%\maintextlabel{fig5}{5}
%\maintextlabel{fig6}{6}
%\maintextlabel{fig7}{7}
%\maintextlabel{fig8}{8}
%\maintextlabel{fig9}{9}
%\maintextlabel{eq_TNM_resp}{1}
%\maintextlabel{eq_gamma_T_Tf}{5}
%\maintextlabel{eq_M}{4}

\begin{center}
   
\large \textbf{Supplementary materials}
\end{center}

\section{Experimental setup.} 

The experimental setup is shown in fig.~\ref{fig1} of the main text. A layer of glycerol (Sigma-Aldrich, $\geq 99.5$~\%) of thickness $h$ and surface $S_\mathrm{glycerol}$ is sandwiched between two electrodes separated by a 13~µm thick PET spacer (Goodfellow). The exact thickness $h = 15 \pm 1~\upmu$m is discussed and justified below. The electrodes are $7\times 15$~mm large and 1~mm thick glass plates covered by a 80~nm thick conductive layer of indium tin oxide (ITO). Two parallel 400~nm thick copper layers with 3~mm width are deposited on each end of the electrodes. This allows the electrical contact with wrapped copper wires maintained with a silver-based conductive resin (EpoTek H20E). Each electrode has an electrical resistance $r=47\pm 2~\Omega$. The ensemble is pressed using a spring against a copper frame, whose temperature is monitored using a Cernox RTD sensor. Surfaces are coated with grease to ensure good thermal contact. The setup is mounted inside a closed copper cell in a cryostat. A 50~W electrical resistance in contact with the cell allows to regulate the sample temperature within 10~mK. A Lakeshore 335 is used to measure the temperature and to control the output power while a proportional integral regulation is assured by a python program running on a computer which allows to interrupt the regulation during 30~s after the heating phase.

\begin{figure}[htbp]
 
\centering
 
\includegraphics[width=6cm]{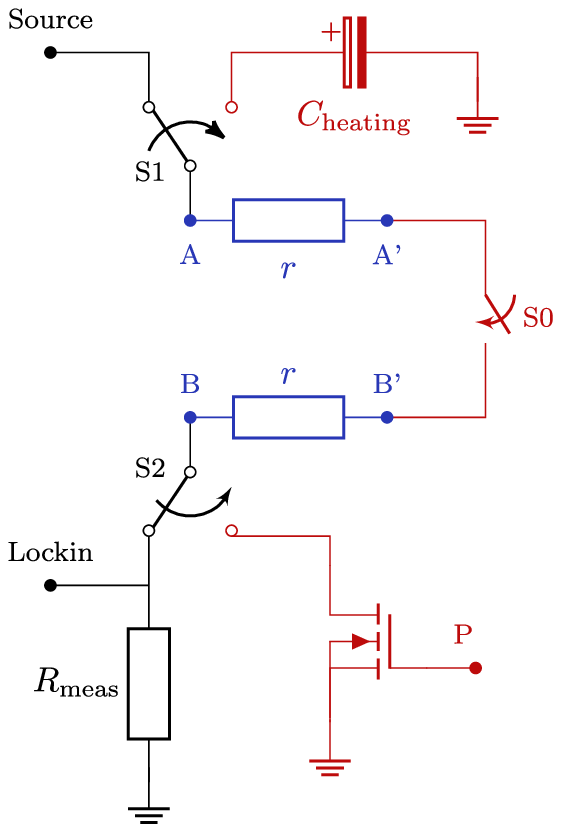}
 \caption{Electrical diagram of the setup. The two electrodes sandwiching the glycerol are shown in blue (A-A' and B-B'). The switches S0, S1, and S2 allow to select between the measurement mode (active here) and the heating mode. The measurement and heating circuit are shown in black and red respectively. The transistor P is used to control the heating pulse sequence described in fig.~\ref{fig2}a.}
 
\label{fig_suppl_schema_elec}
\end{figure}

The electric diagram of the setup is shown in fig.~\ref{fig_suppl_schema_elec}. Switches S0, S1, S2 controlled by an Arduino Leonardo micro-controller allow to select between a measurement mode and a heating mode. 

\textbf{Measurement mode.} In this configuration, the two electrodes form a parallel plate capacitor of admittance $\underline{Y}$ with glycerol and PET spacer as dielectric. It is arranged in series with a measuring resistor $R_\mathrm{meas} = 50$~k$\Omega$. A SR830 DSP lock-in amplifier is used as the source of a 5~Vrms sinusoidal voltage at frequency $f$ alimenting the whole, and to measure the voltage across the $R_\mathrm{meas}$ from which $\underline{Y}(\omega)$ can be deduced. The resistance $r$ of the electrodes does not perturb the impedance measurement as it is much smaller than the leakage resistance of the liquid (the electrical potential can thus be considered homogeneous on all the electrode surface). Between the two electrodes of total surface $S_\mathrm{tot}=(4.9\pm0.1)\times 10^{-5}$~m$^2$, an area $S_\mathrm{glycerol} = (2.7\pm 0.1)\times 10^{-5}$~m$^2$ (measured using a binocular) is occupied by the glycerol while the rest is the PET spacer. The admittance of the system can thus be modeled by:
\begin{align}
\underline{Y}(f) = j \frac{2\pi f \epsilon_0}{h}[\underline{\epsilon}_\mathrm{r} S_\mathrm{glycerol} + \epsilon_\mathrm{r, PET} (S_\mathrm{tot}-S_\mathrm{glycerol})]
\end{align}
where $\epsilon_0$ is the vacuum permittivity,- $\underline{\epsilon}_\mathrm{r} = \epsilon^\prime + j \epsilon^{\prime\prime}$ is the complex relative permittivity of glycerol and $\epsilon_\mathrm{r, PET} = 3.4$ is the relative permittivity of PET (assumed real)~\cite{Polymer_database_Dielectric}. This quantity was measured at $T=300$~K between 1~kHz and 10~kHz: $Im[\underline{Y}/(2\pi f)] = (7.03 \pm 0.06) \times 10^{-10}$~F. Using a tabulated value~\cite{davidson1951dielectric} of $\epsilon^\prime(300~\mathrm{K}) = 42.1$, this leads to $h=15 \pm 1$~µm slightly thicker than the PET spacer thickness which could be explained by the presence of dust or by the existence of a micrometric glycerol layer present between the electrode and the spacer. This thickness value is used in the main text and in the following to convert the measured values of $\underline{Y}$ to values of $\epsilon^\prime$ and $\epsilon^{\prime\prime}$.

\begin{figure}[htbp]
 
\centering
 
\includegraphics[width=16cm]{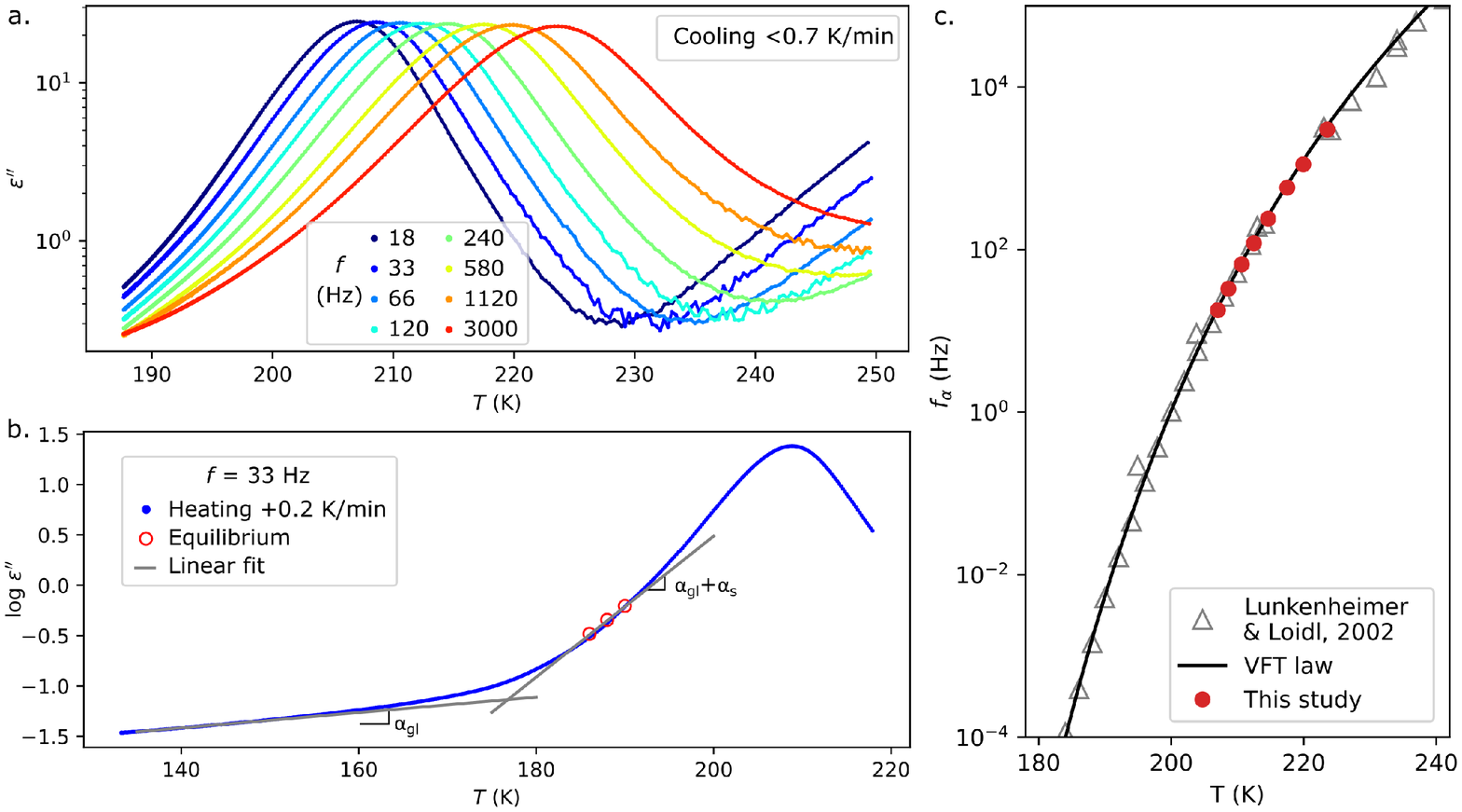}
 \caption{(a) Dielectric loss $\epsilon^{\prime\prime}$  of the glycerol as a function of the temperature $T$ for different frequencies $f$, measured during the cooling of the sample (at a rate $\approx -0.7$~K/min for $T>200$~K and $\approx 0.05$~K/min below, the error on the temperature measurement is smaller than the marker size). (b) Same quantity, measured at $f= 33$~Hz, during heating at 0.2~K/min (blue dots) and measured at equilibrium (red circles). (c) Frequency $f_\alpha$ of the alpha relaxation as a function of the temperature measured in ref.~\citep{lunkenheimer_dielectric_2002} (grey triangles) and in this study (red circles) from the temperature at which the peak is reached in (b).}
  \label{fig_suppl_dielectrique}
\end{figure}

The temperature dependence of $\epsilon^{\prime\prime}$ was measured during the cooling of the sample and is shown in fig.~\ref{fig_suppl_dielectrique}a. Peaks corresponding to the alpha relaxation are well visible for each frequency. The temperature at which they occur is reported for each frequency in fig.~\ref{fig_suppl_dielectrique}c alongside with the alpha relaxation frequency $f_\alpha$ measured by Lunkenheimer and Loidl~\cite{lunkenheimer_dielectric_2002}. The excellent agreement allows to check that the glycerol was not contaminated with water (only a few percent water leads to a significant shift in $f_\alpha$~\cite{ryabov2003features}).

\textbf{Heating mode.} The resistance $r$ of the electrodes can be exploited to heat the liquid with a high heating rate. In this mode, the electrodes form a serial association of resistances (by connecting A' and B') through which an electric current $I(t)$ can flow. This current is delivered by a capacitor ($C_\mathrm{heating} = 530$~µF) previously charged at a voltage $U$ (from 50 to 200~V). The discharge is controlled by a Mosfet transistor and an Arduino micro-controller. Initially, the whole sample is at thermal equilibrium with the cell at temperature $T_0$. The heating process occurs in 3 steps (visible in fig.~\ref{fig2}a of the main text and in fig.~\ref{fig_suppl_Tt}): at $t=0$, the transistor is turned on during a time $t_\mathrm{rise}$ in which the liquid temperature rises and reach $T_0+\Delta T$. At this point, the transistor is actuated during a time $t_\mathrm{up}$ following a prerecorded sequence designed to keep the liquid temperature in an interval $[T_0+\Delta T - \delta T ; T_0+\Delta T]$. Finally, the transistor is turned off and the circuit is switched to measurement mode in approximately 20~ms.

\section{Conversion of the dielectric loss into fictive temperature}

In order to convert the dielectric loss into a fictive temperature, $\log \epsilon^{\prime\prime}(f= 33~\mathrm{Hz})$ was measured during a slow heating ramp (at 0.2 K/min) from 135 to 220 K as shown in fig.~\ref{fig_suppl_dielectrique}b. For temperature significantly smaller than $T_\mathrm{g} = 187$~K no structural relaxation occurs ($T_\mathrm{f}$ is constant) and the variation of the dielectric loss is only due to isotructural changes. This effect is assumed linear with slope $\alpha_\mathrm{gl} = \frac{\Delta \log \epsilon^{\prime\prime}}{\Delta T}|_\mathrm{isostructural} = 0.0075 \pm 0.0005$. In the range 186-192~K, the heating rate is low enough so that the transformation is quasistatic ($T = T_\mathrm{f}$). This is verified by plotting in fig.~\ref{fig_suppl_dielectrique}b the value of $\log \epsilon^{\prime\prime}$ measured equilibrium (red circles). This range being small, the evolution can be assumed linear in first approximation with slope $\alpha_\mathrm{gl}+\alpha_\mathrm{s} = \frac{\Delta \log \epsilon^{\prime\prime}}{\Delta T}|_\mathrm{quasistatic} = 0.070 \pm 0.002$ at $f=33$~Hz. From this, the variation of fictive temperature at constant temperature is obtained by $\Delta T_\mathrm{f} = \frac{\Delta \log \epsilon^{\prime\prime}}{\alpha_\mathrm{s}}$ with $\alpha_\mathrm{s} = 0.063 \pm 0.002$. The fictive temperature obtained from the experimental measurements does not exceed 192~K which justifies the linear approximation.

\section{Thermal numerical simulation}

\subsection{Details and parameters}
The temperature evolution of the liquid during a heating-cooling event is determined from a 1D numerical simulation along the direction orthogonal to the electrodes. This 1D assumption is justified by the small aspect ratio ($\approx 10^{-3}$) of the sandwiched liquid layer. The modelization of the sample geometry is as follows: A region of thickness $h = 15$~µm corresponding to glycerol is surrounded on each side by two regions of thickness $H = 1$~mm corresponding to the glass electrodes. A thermal flux $j(t) = J(t)/2 = \frac{U_\mathrm{h}^2}{2r S_\mathrm{h}} \frac{1}{2}$ (with $S_\mathrm{h} = (6.3\pm0.1)\times 10^{-5}$~m$^2$) can be injected at each of the two boundaries between the domains in order to mimic the Ohmic heating occurring in the resistive layers present at the surface of the electrodes. At the external boundaries of the glass domain, a constant temperature $T_0$ is imposed which limits the duration of the thermal simulation to $t_\mathrm{max} = 350$~ms during which the result does not depend on $H$ (as $\sqrt{t_\mathrm{max}\kappa_\mathrm{glass}/\mathcal{C}_\mathrm{glass}} \ll H$). The thermal parameters needed in the simulation are the thermal conductivity $\kappa$ and the volumetric heat capacity $\mathcal{C}$ for the glass electrode~\cite{Astrath_2005} and for the glycerol~\cite{PhysRevB.34.1631} :
\begin{itemize}
   
\item $\kappa_\mathrm{glass} = 0.93$~W$\cdot$m$^{-1}\cdot$K$^{-1}$ and $\mathcal{C}_\mathrm{p~glass} = 1.6 \times 10^6$~J$\cdot$K$^{-1}\cdot$m$^{-3}$ at 210 K.
   
\item $\kappa_\mathrm{glycerol} = 0.36$~W$\cdot$m$^{-1}\cdot$K$^{-1}$ and $\mathcal{C}_\mathrm{p~glycerol} = 1.4 \times 10^6$~J$\cdot$K$^{-1}\cdot$m$^{-3}$. These values are the one of the glassy state at 190~K, \textit{i.e.} measured at a frequency higher than the relaxation rate~\cite{PhysRevB.34.1631}. The heat capacity doubles in the liquid state due to the higher number of available degrees of freedom. $\mathcal{C}_\mathrm{p~glycerol}$ has an important effect on the temperature evolution during the first millisecond, during which the system is very out of equilibrium in all our experiments. This is why we chose to use the glassy state value. We checked that switching to the liquid value after more than 1~ms does not affect the later temperature evolution.
\end{itemize}

The thermal contact resistance between glycerol and a metal electrode has been shown to be not negligible for such thin samples. From the measurements of Minakov~\textit{et al.}~\cite{minakov2003advanced}, it is possible to extract the contact resistance for glycerol: $r_\mathrm{c} = 4.6 \times 10^{-5}$~m$^2\cdot$K$\cdot$W$^{-1}$ (from the 0.95 multiplicative coefficient on the temperature given by the authors and the thermal resistance of their 250~$\upmu$m thick glycerol sample). For our own sample, this contact resistance is of the same order of the glycerol layer thermal resistance $r_\mathrm{gly} = h/\kappa_\mathrm{glycerol}$. We took this effect into account in our numerical simulation by reducing the thermal conductivity of glycerol in the first 0.1~$\upmu$m near the electrode so that it corresponds to a thermal resistance $r_\mathrm{c}$.

The temperature evolution over a longer time period than 350~ms is estimated by a 3$^\mathrm{rd}$ order polynomial interpolation in $\log t$ (by constraining the continuity of $T$ and its first derivative and its monotonous relaxation to $T_0$ at a time at which the first and second derivative vanish). This does not require any adjustable parameters.

\begin{figure}[htbp]
 
\centering
 
\includegraphics[width=14cm]{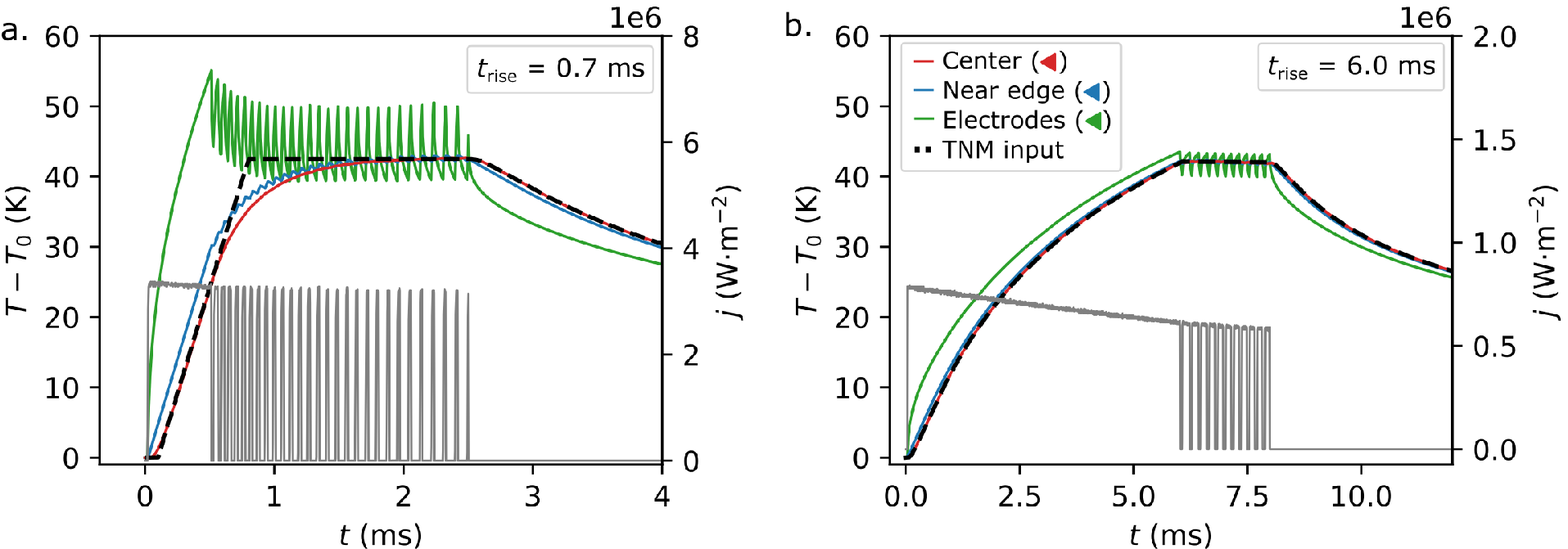}
 \caption{Temperature evolution and heat flux applied to the sample for $t_\mathrm{rise} = 0.7$~ms (a) and 6.0~ms (b). The legend is the same than in fig. 2a of the main text.}
\label{fig_suppl_Tt}
\end{figure}

\subsection{Uncertainties on the temperature evolution}

In order to quantify the uncertainty on the temperature evolution, and more especially on $\Delta T$, it is necessary to evaluate the uncertainties on each parameter of the thermal model.

\begin{figure}[htbp]
 
\centering
 
\includegraphics[width=6cm]{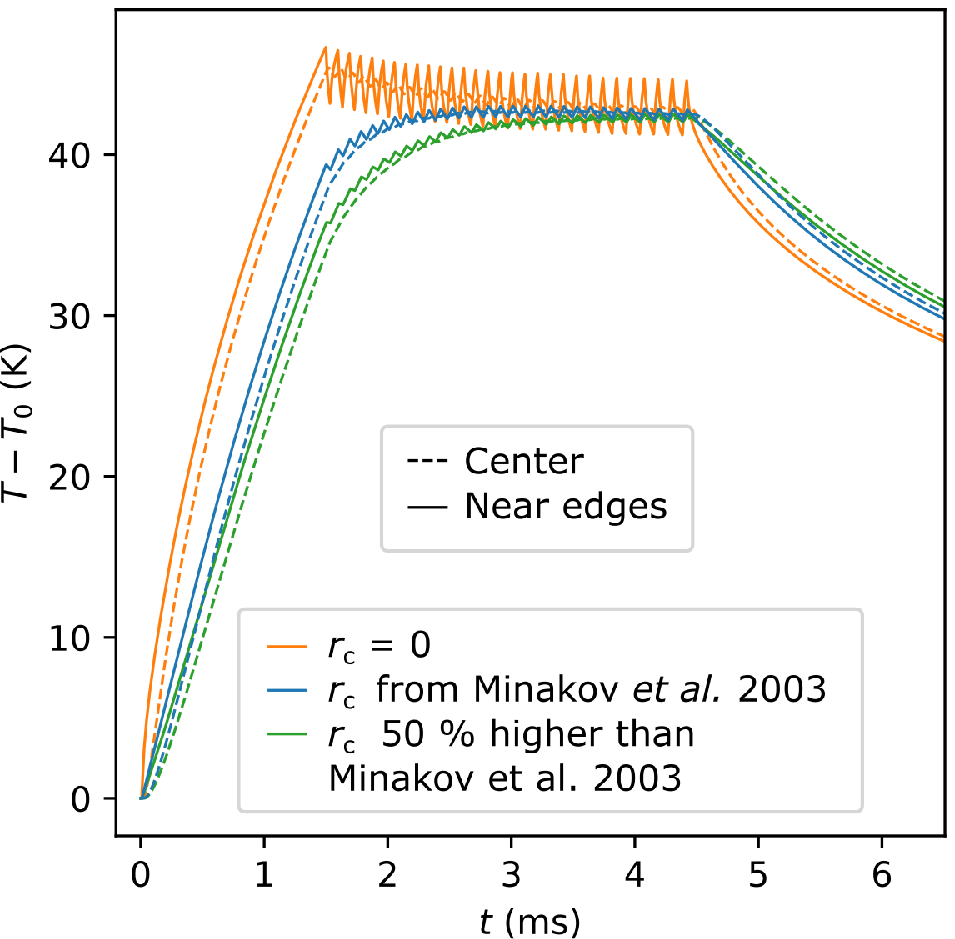}
 \caption{Effect of the thermal contact resistance $r_\mathrm{c}$ between glycerol and electrodes, on the temperature evolution for $t_\mathrm{rise} = 1.5$~ms. The value determined from ref.~\cite{minakov2003advanced} (blue) is compared to the no contact resistance case (orange) and to a resistance 50~\% higher (green).}
\label{fig_suppl_Tt_rc}
\end{figure}

\begin{itemize}
    \item The heat flux $j(t)$ is known from the resistance $r$, the electrode surface $S_\mathrm{h}$ and the voltage $U_\mathrm{h}$. The final uncertainty on $j(t)$ is $\approx 1.8$~\% which leads to the same uncertainty on $\Delta T$.
    \item The glycerol layer thickness $h$ has a $\pm 1$~$\upmu$m uncertainty corresponding to $\pm 0.6$~K on $\Delta T$.
    \item It is not easy to estimate the uncertainty associated with the thermal contact resistance $r_\mathrm{c}$. Fig.~\ref{fig_suppl_Tt_rc} shows the temperature evolution with for the $r_\mathrm{c}$ value taken from ref.~\cite{minakov2003advanced}, for a 50~\% higher value and for $r_\mathrm{c} = 0$. It appears that $r_\mathrm{c}$ has an effect on the temperature evolution during the first millisecond after $t_\mathrm{rise}$ but does not affect much $\Delta T$ at longer times for which the uncertainty is not higher than $\pm 0.5$~K.
\end{itemize}

From this, the total relative uncertainty on $\Delta T$ can be estimated to be of the order of 2.7~\% which corresponds to $\pm 1.2$~K for $\Delta T = 45$~K. 

\section{Application of the TNM model}

\subsection{Parameters from Roed~\textit{et al.}~\cite{roed2019generalized}}
Roed~\textit{et al.}~\cite{roed2019generalized} performed aging experiments on glycerol by measuring the dielectric loss and using an experimental setup~\cite{hecksher2010physical} able to apply ideal temperature steps (with $t_\mathrm{rise}\ll \tau_\alpha$). They successfully fitted all their data with the TNM model by using as a memory kernel $M(\xi)$ the result of one experiment (for ideal steps, eq.~\ref{eq_TNM_resp} of the main text simplifies in $T_\mathrm{f}(\xi) = T(0) + \Delta T[1-M(\xi)]$) and the following expression for the relaxation rate:
\begin{align}
\gamma(t) = \gamma_\mathrm{eq} \exp\left( \frac{X(t)-X_\infty}{X_\mathrm{const}}\right)  
\label{eq_suppl_gamma_Roed}
\end{align}
where $X(t) = \log \epsilon^{\prime \prime} (t)$, $X_\infty$ is the equilibrium value reached after the experiment and $X_\mathrm{const}=0.16$ is the only adjustable parameter.

If the formalism may appear a bit different that the one that we used in our work, it is in fact fully compatible. For a step experiment from $T_0$ to $T_\infty$ at $t=0= \xi$,  the measured quantity $X$ can be converted into a fictive temperature in the same way as in our work: 
\begin{align}
X(t) - X_\infty = \alpha_\mathrm{s}(T_\mathrm{f}(t) - T_\infty)
\end{align}
as $T=T_\mathrm{\infty}$ is constant for $t>0$. The value of $\alpha_\mathrm{s}$ in Roed experiments can be deduced from their fig.~2a ($\Delta X_\mathrm{eq} = (\alpha_\mathrm{gl}+\alpha_\mathrm{s}) \Delta T$, with $\alpha_\mathrm{gl}+\alpha_\mathrm{s} = 0.151 \pm 0.005$) and from their fig.~3 from which they plot:
\begin{align}
R(t) = \frac{X(t)-X_\infty}{X_0-X_\infty} = \frac{\alpha_\mathrm{s}(T_\mathrm{f}(t)-T_\infty)}{(\alpha_\mathrm{gl}+\alpha_\mathrm{s})(T_0-X_\infty)}
\end{align}
From the initial value $R(0) = 0.65 \pm 0.05 = \alpha_\mathrm{s}/(\alpha_\mathrm{gl}+\alpha_\mathrm{s})$, we can deduce $\alpha_\mathrm{s} = 0.10 \pm 0.01$

In the range 176-184~K in which they work, $\log \tau_\alpha$ can be linearized in $\ln \tau_\alpha(T) = -AT + B$ where $A = 0.76 \pm 0.04$ can be deduced from their fig.~4 and $B$ an uninteresting constant. It is worth noting that their measured values of $\tau_\alpha$ are systematically smaller than those of Lunkenheimer\& Loidl~\cite{lunkenheimer_dielectric_2002} and ours by approximately a decade. This may be due to the presence of a very small amount of water in the samples. For a constant $T=T_\infty$ and a small $T_\mathrm{f}-T_\infty$, the expression of $\gamma(T, T_\mathrm{f})$ given in eq.~\ref{eq_gamma_T_Tf} of the main text, can be simplified in:

\begin{align}
\gamma(t) &= \gamma^x_\mathrm{eq}(T_\infty) \gamma^{1-x}_\mathrm{eq}(T_\mathrm{f}(t))\\
&= e^{x (AT_\infty+ B) + (1-x)(AT_\mathrm{f}(t)+ B)}\\
&= \gamma_\mathrm{eq}(T_\infty) \exp \left( \frac{A(1-x)(X(t)-X_\infty)}{\alpha_\mathrm{s}}\right)
\end{align}
By identification with eq.~\ref{eq_suppl_gamma_Roed}, this lead to $x = 1 - \alpha_\mathrm{s}/(X_\mathrm{const}A) = 0.18 \pm 0.09$.

\medskip

\begin{figure}[htbp]
\centering
\includegraphics[width=18cm]{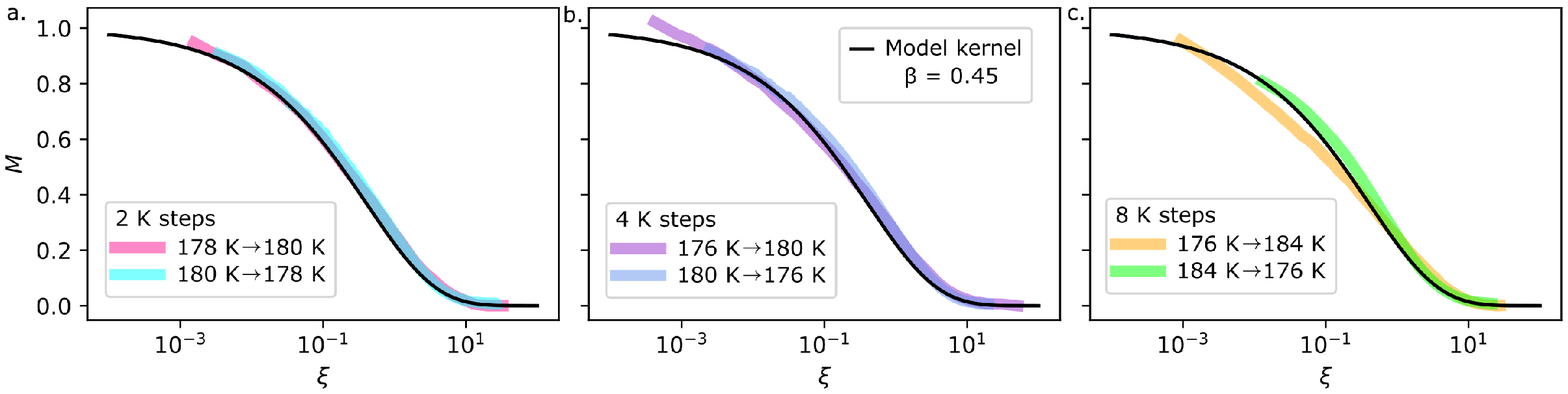}
 \caption{Memory kernel obtained from the experimental data from Roed~\textit{et al.}~\cite{roed2019generalized} and their single parameter $X_\mathrm{const} = 0.16$ (colored lines) for temperature steps of 2~K (a), 4~K (b) and 8~K (c). The black curves are a fit of eq.~\ref{eq_M} of the main text with $\beta = 0.45$.}
\label{fig_suppl_roed}
\end{figure}

The authors chose to not directly extract the memory kernel $M(\xi)$ but rather to use it as a calculation intermediate thus removing the need for any assumption on its form. For the purpose of comparison with the present work, it is interesting to extract this memory kernel by computing for each step experiments:
\begin{align}
    M(\xi) = R(\xi) \frac{\alpha_\mathrm{gl} + \alpha_\mathrm{s}}{\alpha_\mathrm{s}}, \quad \xi(t) = \int_0^t \gamma(t')dt'
\end{align}
Fig.~\ref{fig_suppl_roed} shows $M(\xi)$ for upwards and downwards temperature steps of amplitude ranging from 2 to 8~K. All these data could be fitted with the stretched exponential of eq.~\ref{eq_M} of the main text with $\beta = 0.45$  which corresponds to $\xi_0 = 0.4$.

\subsection{Cooling phase}
\begin{figure}[htbp]
\centering
\includegraphics[width=13cm]{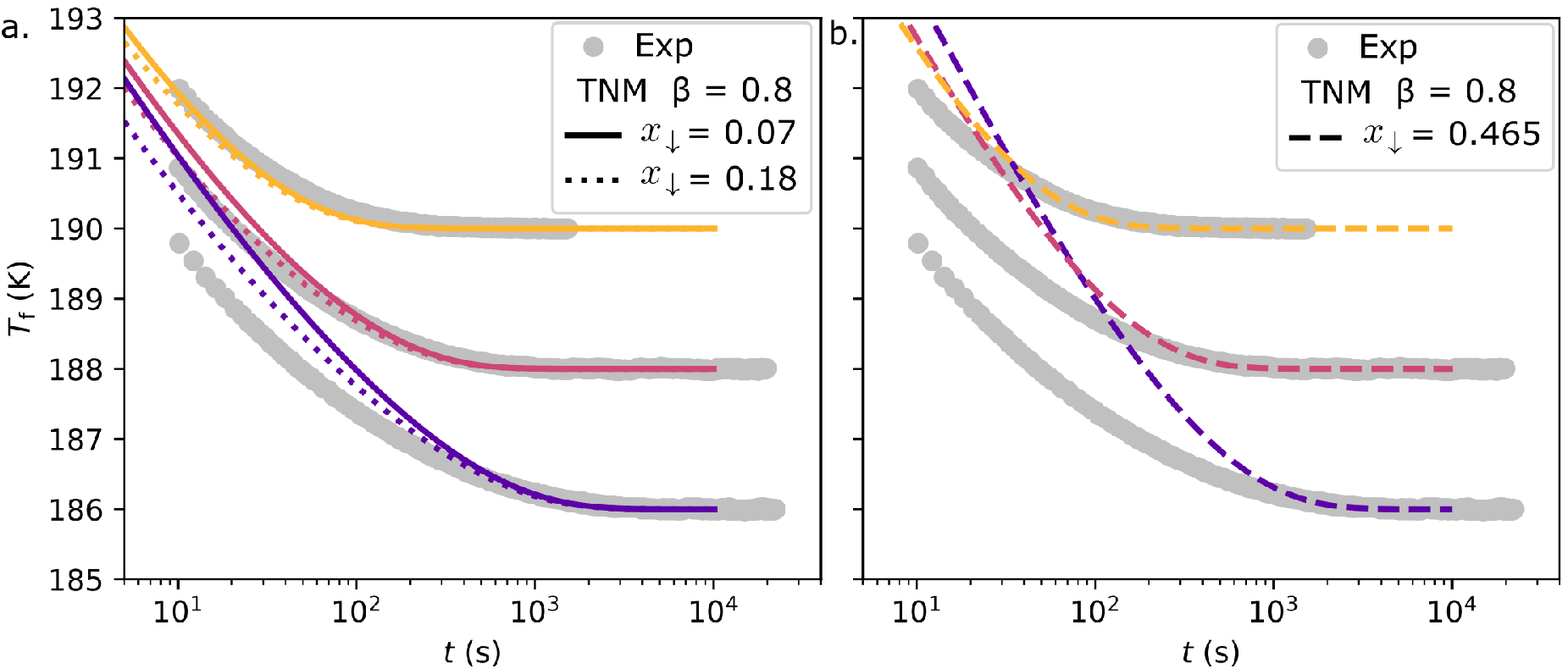}
 \caption{Comparison of the experimental data (grey markers) and the TNM model (lines with $\beta = 0.8$, $x_{\downarrow}$ in the legend) in the cooling phase starting from an equilibrated state at $T_0 + \Delta T$ (with $\Delta T = 41$~K).}
\label{fig_suppl_desc}
\end{figure}

Fig.~\ref{fig_suppl_desc} compares the relaxation of $T_\mathrm{f}$ measured during the cooling from $\epsilon_\mathrm{max}^{\prime\prime}$ with the prediction of the TNM model with $\beta = 0.8$ and various values of $x_{\downarrow} = 0.07$, 0.18 (a) and 0.465 (b).

\subsection{Idealized or realistic temperature profile}

\begin{figure}[htbp]
\centering
\includegraphics[width=13cm]{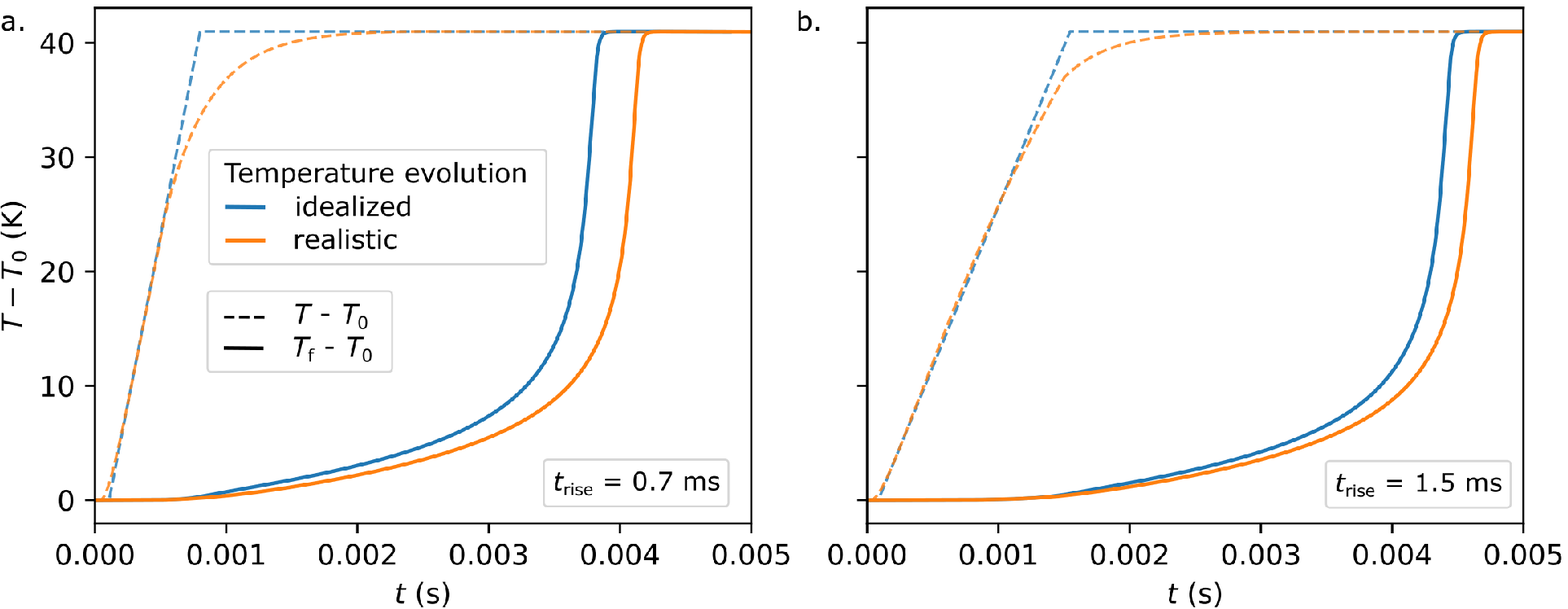}
 \caption{Effect of the temperature evolution $T(t)$ (dashed lines) on the response $T_\mathrm{f}$ (solid lines) computed using the TNM model (with parameters $\beta = 0.8$ and $x_{\uparrow} = 0.465$) for $T_0 = 188$~K, $\Delta T = 41$~K, $t_\mathrm{rise} = 0.7$~ms (a) and 1.5~ms (b). The realistic temperature evolution undergone by the liquid during the experiments (orange) is compared to the idealized version (blue) used in the main text.}
\label{fig_suppl_TTf_idealized}
\end{figure}

The temperature profiles shown in fig.~2a of the main text ($t_\mathrm{rise} = 1.5$~ms) and in fig.~\ref{fig_suppl_Tt}a of the supplementary ($t_\mathrm{rise} = 0.7$~ms) both display, after a time $t_\mathrm{rise}$, a slower increase, lasting for $0.5-1$~ms, before stabilizing at $T_0 + \Delta T$. In order to keep the analysis as simple as possible, we used, in the main text, an idealized version of $T(t)$ in which $T = T_0 + \Delta T$ after $t_\mathrm{rise}$ (see dashed blue lines in fig.~\ref{fig_suppl_TTf_idealized}). We quantify here the error caused by this approximation. The response $T_\mathrm{f}(t)$ computed with the idealized profile as input (solid blue lines) is compared to the one that would be obtained using a more realistic temperature evolution (orange lines). The latter responses are ed toward longer times as a consequence of the smaller relaxation rate $\gamma(T, T_\mathrm{f})$ during the first millisecond after $t_\mathrm{rise}$. The time needed to equilibrate the liquid at $T_0+\Delta T$ is increased by 8.5~\% for $t_\mathrm{rise} = 0.7$~ms and 4.5~\% for $t_\mathrm{rise} = 1.5$~ms.

\subsection{Heating cooling experiments}
\begin{figure}[htbp]
\centering
\includegraphics[width=13cm]{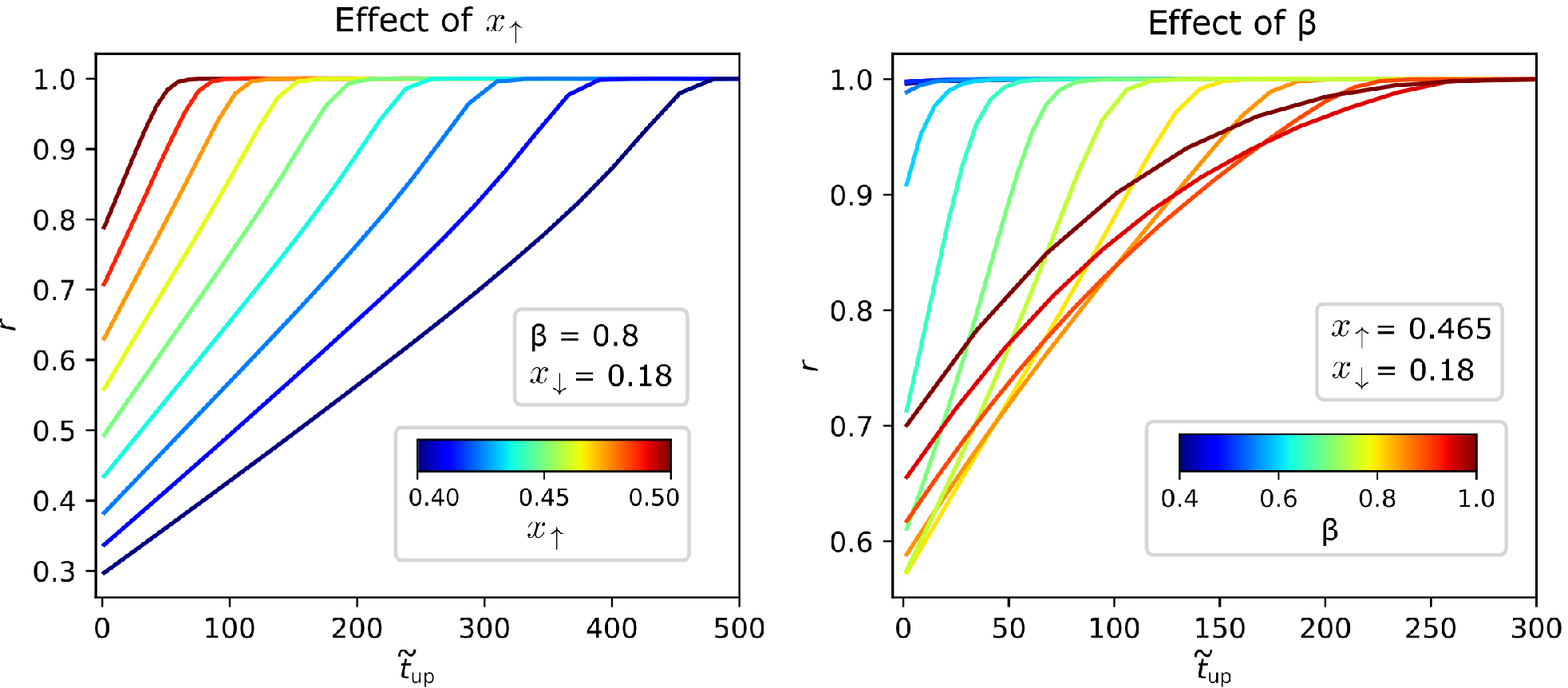}
 \caption{Influence of the free parameters $\beta$ and $x_{\uparrow}$ on the result of heating-cooling experiments for $t_\mathrm{rise} = 1.5$~ms, $T_0 = 188$~K, $\Delta T = 41$~K and $x_{\downarrow} = 0.18$.}
\label{suppl_tnm_beta_x}
\end{figure}

See fig.~\ref{suppl_tnm_beta_x} for the effect of $\beta$ and $x_{\uparrow}$ on the $r(\tilde{t}_\mathrm{up})$ prediction of the TNM model..

\subsection{Correspondence between $R$ and $r$}
\begin{figure}[htbp]
 
\centering
 
\includegraphics[width=10cm]{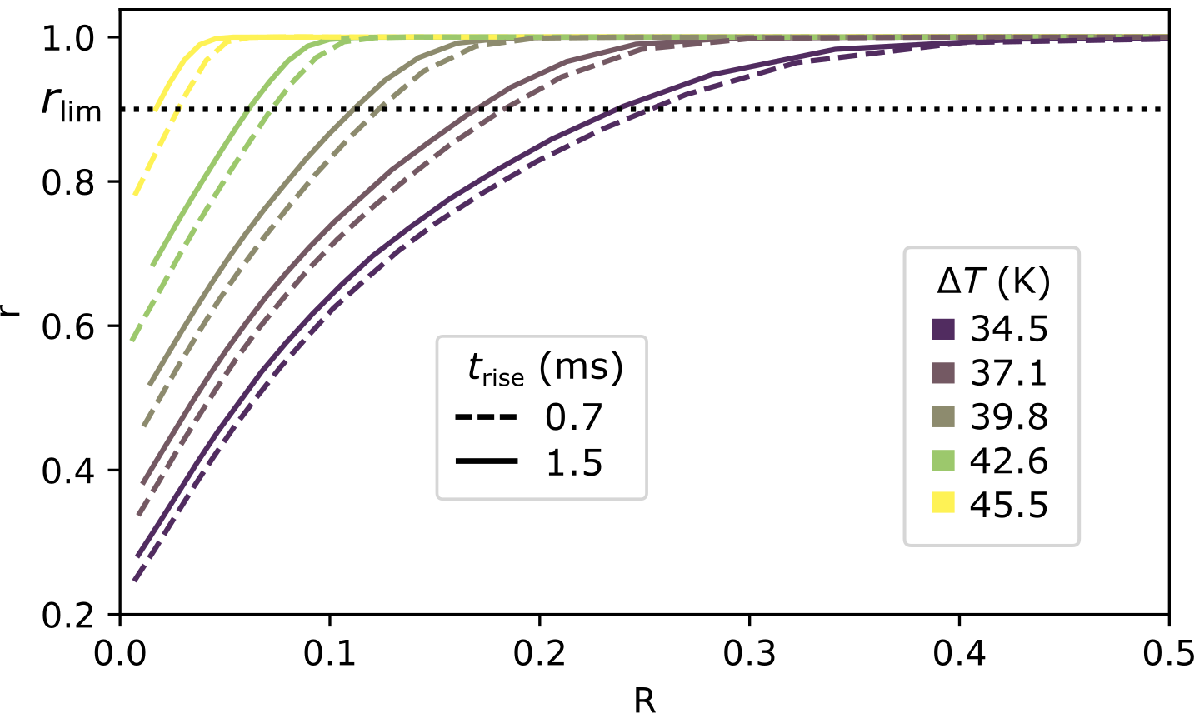}
 \caption{Correspondence between the ratio $r$ measured from the experiments when starting the cooling phase after $t_\mathrm{up}$ and the corresponding response $R = (T_\mathrm{f}(t_\mathrm{up})-T_0)/\Delta T$ of the liquid to the heating phase. This was determined from the application of the TNM model at $T_0 = 188$~K (with parameters $\beta=0.8$, $x_{\uparrow} = x_\mathrm{L} = 0.465$ and $x_{\downarrow} = x_\mathrm{S} = 0.18$).}
 
\label{suppl_Rr}
\end{figure}

See fig.~\ref{suppl_Rr}.

\end{document}